\documentclass[12pt]{iopart}

\begin{document}

\title[The problem of motion in gauge theories of gravity]{The problem of motion in gauge theories of gravity}

\author{Serhii Samokhvalov}

\address{Dniprovsk State Technical University, Kamianske, UKRAINE}
\ead{serh.samokhval@gmail.com}
\vspace{10pt}
\begin{indented}
\item[]August 2020
\end{indented}

\begin{abstract}
In this article we consider the problem to what extent the motion of gauge-charged matter that generates the gravitational field can be arbitrary, as well as what equations are superimposed on the gauge field due to conditions of compatibility of gravitational field equations. Considered problem is analyzed from the point of view symmetry of the theory with respect to the generalized gauge deformed groups without specification of Lagrangians.

In particular it is shown, that the motion of uncharged particles along geodesics of Riemannian space is inherent in an extremely wide range of theories of gravity and is a consequence of the gauge translational invariance of these theories under the condition of fulfilling equations of gravitational field. In the cause of gauge-charged particles, the Lorentz force, generalized for gauge-charged matter, appears in equations of motion as a consequence of the gauge symmetry of the theory under the condition of fulfilling equations of gravitational and gauge fields. In addition, we found relationships of equations for some fields that follow from the assumption about fulfilling of equations for other fields, for example, relationships of equations of the gravitational field and the gauge field of internal symmetry which follow from the assumption about fulfilling of equations of matter fields. In particular, we obtained the identity that generalizes in the case of arbitrary gauge field (and in the presence of gauge-charged matter) the identity found by Hilbert for the electromagnetic field.

At the end of the article there is an Appendix, which briefly describes the main provisions and facts from the theory of generalized gauge deformed groups and presents the main ideas of a single group-theoretical interpretation of gauge fields of both external (space-time) and internal symmetry, which is an alternative to their geometric interpretation.

\end{abstract}

\vspace{2pc}
\noindent{\it Keywords}: gauge theories of gravity, generalized gauge deformed groups, problem of motion
%\maketitle

\section{Introduction}

One of the most striking features of the general relativity (GR) is the fact that the matter that generates gravitational field cannot move arbitrarily, but must obey certain equations that follow from equations of the gravitational field as a condition of their compatibility. This fact was first noted in the fundamental Hilbert's work \cite{Hilbert}, in which equations of GR saw the world for the first time as variational Lagrange equations. Hilbert showed that in the case when fulfilling equations of the gravitational field which were born by an electromagnetic field, four linear combinations of equations of the electromagnetic field and their derivatives are zero due to the general covariance of the theory. It is known that this is what stimulated E. Noether to invent her famous theorem \cite{Noether}.

As for "solid matter", for the compatibility of equations of the gravitational field, it is necessary that particles of dust matter move along geodesics of Riemannian space, which describes the gravitational field. This fact was pointed out in the work of A. Einstein and J. Grommer \cite{Einstein27} and according to V. Fock \cite{Fock} it is one of the main justifications of GR (although it should be noted that even before the creation of GR it was known that the motion along geodesics is a consequence of the condition of covariant conservation of energy-momentum of matter \cite{Einstein13}). This remarkable feature of GR all his life inspired Einstein to search on the basis of GR such theory from which it would be possible to derive all fundamental physics, including quantum mechanics.

Interest in this problem (following Einstein, we name it the problem of motion) has resumed in our time in connection with the registration of gravitational waves and analysis of the conditions of their radiation, i.e. the need for its direct application in gravitational-wave astronomy \cite{Oltean}.

In article \cite{Sam}, the problem was considered to what extent the motion of electrically charged matter that generates the gravitational field can be arbitrary, as well as what equations are superimposed on the electromagnetic field due to the conditions of compatibility of equations of gravitational field in this case. The study was grounded on group-theoretical analysis based on the symmetry of the theory of relatively generalized gauge deformed group $G_{M}^{g} =U(1)^{g} \left. \times \right)\, T_{M}^{g} $ \cite{TMF}, which is a semidirect product of the gauge group of electrodynamics $U(1)^{g} $ and the generalized gauge group of translations $T_{M}^{g} $. In the aforementioned article, in particular, it was shown that due to the $G_{M}^{g} $-invariance of the theory, when fulfilling equations of the gravitational field: a) in the case of dust matter, particles move along geodesics, and in the presence of an electromagnetic field (if the particles are charged) they are additionally affected by the Lorentz force; b) provided that the electric charge and the total energy-momentum of the system of gravitational and electromagnetic fields and the field of charged matter are conserved, regardless of the method of parameterization of group $G_{M}^{g} $, the electromagnetic field obeys Maxwell's equations.

The results of article \cite{Sam} were determined solely by the symmetry of the theory and did not depend on the specific type of Lagrangians of both gravitational and electromagnetic fields. The only limitation was that it was assumed that Lagrangians depend on derivatives of field variables not higher than the first order. Thus, the results obtained in \cite{Sam} are valid not only in GR, but also in a number of other theories of gravity, in particular in the currently popular $f(T)$-theories, which can provide a theoretical interpretation of the late-time universe acceleration (alternative to a cosmological constant), avoidance of the Big Bang singularity etc. \cite{Cai}.

This article generalizes the results of article \cite{Sam} in the following directions. Firstly, the restrictions on the order of derivatives of field variables in the Lagrangian of the gravitational field are removed. Secondly, instead of the electromagnetic field, arbitrary gauge field of internal symmetry $V^{g} $ are considered. This will allows one to analyze the problem of motion, in addition to GR and $f(T)$-theories, also in a fairly wide class of theories of gravity, which are now intensively studied, in particular, such as $f(R)$-theories \cite{Pintoa}, Lovelock gravity \cite{Petrov}, and others which do not have problems with local-Lorentz invariance, as $f(T)$-theories \cite{Maluf}. In addition, this analysis can be performed in the case when the sources of the gravitational field are arbitrary gauge-charged matter fields and corresponding gauge fields of internal symmetry (and not only electrically charged matter and electromagnetic field). Models where non-Abelian gauge vector fields are the unique source of inflation and dark energy are now intensively studied \cite{Guarnizo}.

This paper is structured as follows.

In Section 2 we prove general Theorem about quasilocality of charges associated with generalized gauge symmetries $G_{M}^{g} $ for theories with Lagrangians that depend on fields and their derivatives up to an arbitrary order.

In Section 3 we study the problem of motion for the abovementioned theories of gravity in terms of their symmetry with respect to the generalized gauge translations group $T_{M}^{g} $ \cite{TMF} and show that equations of the gravitational field in an arbitrary gauge theory of group $T_{M}^{g} $ can be written both in the form similar to the form of Einstein equation and in a superpotential form, i.e. in the form of the expression of the total energy-momentum tensor of the gravitational system through the superpotential (in the form similar to the form of Yang-Mills equations). In addition, Proposition 1 is proved, from which, in particular, follows that in an arbitrary gauge theory of generalized gauge translations group $T_{M}^{g} $ for the compatibility of the gravitational field equations dust matter must move along geodesics of Riemannian space corresponding to the solution of the gravitational field equations.

In Section 4 we assume that matter carries gauge charges and so gauge fields of internal symmetry $V^{g} $ are present. The symmetry of the theory in this case extends to group $G_{M}^{g} =V^{g} \left. \times \right)\, T_{M}^{g} $, which is a semidirect product of the groups $V^{g} $ and $T_{M}^{g} $. Here we show that equations of arbitrary gauge field of internal symmetry $V^{g} $ (regardless of the specific type of its Lagrangian) in the presence of gravitational field can be written both in the form of Einstein equation and in superpotential form, i.e. as an expression of the total current of gauge charges of the system through the superpotential (in the form of Yang-Mills equations). In addition, Proposition 2 is proved, in which relationships of equations for some fields that follow from the assumption about fulfilling of equations for other fields are revealed (for example, relationships of equations of the gravitational field and the gauge field of internal symmetry which follow from the assumption about fulfilling of equations of matter fields, etc.). In particular, we obtained the identity that generalizes in the case of arbitrary gauge field (and in the presence of gauge-charged matter) the identity found by Hilbert for the electromagnetic field \cite{Hilbert}.

The main results of this work are based on a group-theoretic analysis of symmetry of theories relatively generalized gauge deformed groups that act non-trivially on the space-time manifold. Since the method of deformations of generalized gauge groups, despite its effectiveness, has not yet become widely known, the Appendix is included, where the basic concepts of this method is presented as transparently as possible and without mathematical details, but with all the necessary references to primary sources. In particular, in the Appendix the concept of the generalized group $T_M^g$  of gauge translations in curved Riemannian space as a deformed group of its diffeomorphisms is discussed (Appendix A.3). The generalized gauge deformed group $G_M^g$ has the structure of a semidirect product of the ordinary gauge group of internal symmetry $V^g$ and the generalized group $T_M^g$ of gauge translations (Appendix A.2). The method of forming a semidirect product of these groups is described. It ensures the fact that the obtained group $G_M^g$ describes translations in a curved space of fiber bundles with connection. The Appendix presents the main ideas of a single group-theoretical description of gauge fields of both external (space-time) and internal symmetry which acquire single interpretation as deformation coefficients of suitable generalized gauge deformed groups that is an alternative to their geometrical interpretation. It should be noted here that group-theoretical description of gauge fields is more fundamental than geometrical description because according to Klein’s Erlangen Program geometrical structures are defined by the groups, which acts on manifolds.

\section{Quasilocality of gauge charges}

Let on the space-time manifold $M$ with coordinates $x^{\mu } $ (for which we will use the Greek indices of the middle of the alphabet) the system of fields $q^{I} (x)$ (the field index is $I$)  and the action

\[
S=\int L(q ,\partial {\kern 1pt} q\, ...{\kern 1pt} \, \partial ^{(n)} q)dx,
\]
are given, where $L$ is the Lagrangian, which depends on the fields and their derivatives up to the $n$-th order, and $dx$ is an element of coordinate volume. Let's consider infinitesimal transformations of coordinates and fields:

\[
x'^{\mu } =x^{\mu } +\delta {\kern 1pt} x^{\mu },\qquad q'^{I} (x)=q^{I} (x)+\delta {\kern 1pt} q^{I},
\]
in which the action acquires addend $\bar{\delta }S=\int \delta 'Ldx $, where $\delta 'L=L'(x')J-L(x)$ is the integral variation of the Lagrangian $L$, and $L'(x')$ means the Lagrangian calculated for the fields $q'$ at the point $x'$, $J$ is the Jacobian of the transformation $x'(x)$. By direct variation we obtain:

\begin{equation}\label{eq1}
\delta 'L=[L]{\kern 1pt} _{I} {\kern 1pt} \delta {\kern 1pt} q^{I} +\partial _{\sigma } \bar{W}^{\sigma } ,\qquad \bar{W}^{\sigma } =W^{\sigma } +L\delta {\kern 1pt} x^{\sigma },
\end{equation}
where

\begin{equation} \label{eq2} [L]{\kern 1pt} _{I} :=\sum _{k=0}^{n}(-1)^{k} \partial _{\sigma _{1} ...\sigma _{k} }^{(k)} \partial {\kern 1pt} _{I}^{\sigma _{1} ...\sigma _{k} } L =\partial {\kern 1pt} _{I}^{} L-\partial _{\sigma }^{} \partial {\kern 1pt} _{I}^{\sigma } L+\partial _{\sigma {\kern 1pt} \rho }^{(2)} \partial {\kern 1pt} _{I}^{\sigma {\kern 1pt} \rho } L-... \end{equation}
are variational derivatives of Lagrangian $L$ with respect to the fields $q^{I} $,

\[
W^{\sigma } :=\sum _{k=0}^{n-1}[L]{\kern 1pt} _{I}^{\nu _{1} ...\nu _{k} {\kern 1pt} \sigma } \partial _{\nu _{1} ...\nu _{k} }^{(k)} \delta {\kern 1pt} q^{I}  =[L]{\kern 1pt} _{I}^{\sigma } \delta {\kern 1pt} q^{I} +[L]{\kern 1pt} _{I}^{\nu {\kern 1pt} \sigma } \partial _{\nu }^{} \delta {\kern 1pt} q^{I} +...
\]
where in turn

\begin{equation} \label{eq3} [L]{\kern 1pt} _{I}^{\nu _{1} ...\nu _{l} } :=\sum _{k=0}^{n-l}(-1)^{k} \partial _{\sigma _{1} ...\sigma _{k} }^{(k)} \partial {\kern 1pt} _{I}^{\nu _{1} ...\nu _{l} {\kern 1pt} {\kern 1pt} \sigma _{1} ...\sigma _{k} } L =\partial {\kern 1pt} _{I}^{\nu _{1} ...\nu _{l} } L-\partial _{\sigma }^{} \partial {\kern 1pt} _{I}^{\nu _{1} ...\nu _{l} \, \sigma } L+... \end{equation}
are variational derivatives of Lagrangian $L$ with respect to the derivations of the fields $\partial _{\nu _{1} ...\nu _{l} }^{(l)} q^{I} $. Notations are used here $\partial _{\sigma}  :=\partial /\partial x^{\sigma} $,  $\partial _{\sigma _{1} ...\sigma _{k} }^{(k)} :=\partial ^{(k)} /\partial x^{\sigma _{1} } ...\, \partial x^{\sigma _{k} } $, $\partial {\kern 1pt} _{I}^{\sigma _{1} ...\sigma _{k} } :=\partial /\partial _{\sigma _{1} ...\sigma _{k} }^{(k)} q^{I} $. Obviously, from (\ref{eq3}) follows (\ref{eq2}) if $l=0$.

Let the infinitesimal action of a generalized gauge group $G_{M}^{g} $ (Appendix) be given by formulas:

\begin{equation}\label{eq4}
\delta x^{\mu } =h_{a}^{\mu } g^{a} , \qquad \delta {\kern 1pt} q^{I} =\sum _{p=0}^{m}a{\kern 1pt} _{\, \, a}^{I{\kern 1pt} {\kern 1pt} \mu _{1} ...\mu _{p} } \partial _{\mu _{1} ...\mu _{p} }^{(p)} g^{a}  =a{\kern 1pt} _{\, a}^{I{\kern 1pt} {\kern 1pt} } g^{a} +a{\kern 1pt} _{\, \, a}^{I{\kern 1pt} {\kern 1pt} \mu } \partial _{\mu }^{} g^{a} +...
\end{equation}
where $h_{a}^{\mu } $ and $a{\kern 1pt} _{\, \, a}^{I{\kern 1pt} {\kern 1pt} \mu _{1} ...\mu _{s} } $ are functions of $x^{\mu } $ and $q^{I} (x)$ which specify this action, and $g^{a} (x)$ are the infinitesimal parameters of the group $G_{M}^{g} $. In this case, the functions $\bar{W}^{\sigma } $ that determine the integral variation of the Lagrangian (\ref{eq1}) are expressed in terms of group parameters and their derivatives:

\begin{equation}\label{eq5}
\bar{W}^{\sigma } =-\sum _{p=0}^{m+n-1}J{\kern 1pt} _{a}^{\mu _{1} ...\mu _{p} {\kern 1pt} \sigma } \partial _{\mu _{1} ...\mu _{p} }^{(p)} g^{a}  ,
\end{equation}
where if $p=0$

\begin{equation}\label{eq6}
J{\kern 1pt} _{a}^{\sigma } :=-\sum _{k=0}^{n-1}{\kern 1pt}  [L]{\kern 1pt} _{I}^{\nu _{1} ...\nu _{k} {\kern 1pt} \sigma } \partial _{\nu _{1} ...\nu _{k} }^{(k)} a{\kern 1pt} _{\, \, a}^{I{\kern 1pt} } -L{\kern 1pt} h_{a}^{\sigma } ,
\end{equation}
and if $p>0$

\begin{equation}\label{eq7}
J{\kern 1pt} _{a}^{\mu _{1} ...\mu _{p} {\kern 1pt} \sigma } :=-\sum _{l=p-m}^{p}\, \sum _{k=l}^{n-1}C_{k}^{\,l} \, {\kern 1pt} [L]{\kern 1pt} _{I}^{ \mu _{1} ...\mu _{l}\, \nu _{l+1} ...\nu _{k} {\kern 1pt} \sigma } \partial _{\nu _{l+1} ...\nu _{k} }^{(k-l)} a{\kern 1pt} _{\, \, a}^{I{\kern 1pt} {\kern 1pt} \mu _{l+1} ...\mu _{p} }   .
\end{equation}
Here $C_{k}^{\,l} $ is the binomial coefficients. The sequence of indices $ \mu _{1} ...\mu _{p} $ in expression (\ref{eq7}) for $ J{\kern 1pt} _{a}^{\mu _{1} ...\mu _{p} {\kern 1pt} \sigma } $ can be arbitrary, but we will not perform their symmetrization, reserving the right further to choose their sequence convenient for us.

For $p=1$ and 2 from (\ref{eq7}) follows:

\[
J{\kern 1pt} _{a}^{\mu {\kern 1pt} {\kern 1pt} \sigma } =-\sum _{k=0}^{n-1}[L]{\kern 1pt} _{I}^{\nu _{1} ...\nu _{k} {\kern 1pt} \sigma } \partial _{\nu _{1} ...\nu _{k} }^{(k)} a{\kern 1pt} _{\, \, a}^{I{\kern 1pt} {\kern 1pt} \mu }  -\sum _{k=1}^{n-1}k{\kern 1pt} \, [L]{\kern 1pt} _{I}^{\mu {\kern 1pt} {\kern 1pt} \nu _{2} ...\nu _{k} {\kern 1pt} \sigma } \partial _{\nu _{2} ...\nu _{k} }^{(k-1)} a{\kern 1pt} _{\, a}^{I{\kern 1pt} {\kern 1pt} },
\]

\[
\eqalign{J{\kern 1pt} _{a}^{\mu {\kern 1pt} \rho {\kern 1pt} \sigma } =-\sum _{k=0}^{n-1}[L]{\kern 1pt} _{I}^{\nu _{1} ...\nu _{k} {\kern 1pt} \sigma } \partial _{\nu _{1} ...\nu _{k} }^{(k)} a{\kern 1pt} _{\, \, a}^{I{\kern 1pt} {\kern 1pt} \mu {\kern 1pt} \rho }  -\sum _{k=1}^{n-1}k{\kern 1pt} \, [L]{\kern 1pt} _{I}^{ \mu {\kern 1pt} {\kern 1pt} \nu _{2} ...\nu _{k} {\kern 1pt} \sigma } \partial _{\nu _{2} ...\nu _{k} }^{(k-1)} a{\kern 1pt} _{\, {\kern 1pt} a}^{I\, {\kern 1pt} {\kern 1pt} \rho } \cr
\qquad \quad \; -\sum _{k=2}^{n-1}\frac{k(k-1)}{2} {\kern 1pt} \, [L]{\kern 1pt} _{I}^{\mu {\kern 1pt} \rho {\kern 1pt} \nu _{3} ...\nu _{k} {\kern 1pt} \sigma } \partial _{\nu _{3} ...\nu _{k} }^{(k-2)} a{\kern 1pt} _{\, a}^{I\, }.}
\]

Directly from definition (\ref{eq5}) follows:

\begin{equation}\label{eq9}
 \partial _{\sigma } \bar{W}^{\sigma } =-\sum _{p=0}^{m+n}S{\kern 1pt} _{a}^{\mu _{1} ...\mu _{p} {\kern 1pt} } \partial _{\mu _{1} ...\mu _{p} }^{(p)} g^{a},
\end{equation}
where

\[
 S{\kern 1pt} _{a}^{\mu _{1} ...\mu _{p} {\kern 1pt} } :=J{\kern 1pt} _{a}^{\mu _{1} ...\mu _{p} {\kern 1pt} } +\partial _{\sigma } J{\kern 1pt} _{a}^{\mu _{1} ...\mu _{p} {\kern 1pt} \sigma }.
\]

Suppose now that action $S$ is invariant with respect to transformations (\ref{eq4}) of the group $G_{M}^{g} $. The condition of $G_{M}^{g} $-symmetry of action $S$ is $\delta 'L=0$, hence, taking into account (\ref{eq1}), it follows $[L]{\kern 1pt} _{I} {\kern 1pt} \delta {\kern 1pt} q^{I} =-\partial _{\sigma } \bar{W}^{\sigma } $. Therefore, taking into account (\ref{eq4}) and (\ref{eq9}), we obtain:

\[
\sum _{p=0}^{m}[L]{\kern 1pt} _{I} {\kern 1pt} a{\kern 1pt} _{\, \, a}^{I{\kern 1pt} {\kern 1pt} \mu _{1} ...\mu _{p} } \partial _{\mu _{1} ...\mu _{p} }^{(p)} g^{a}  =\sum _{p=0}^{m+n}S{\kern 1pt} _{a}^{\mu _{1} ...\mu _{p} {\kern 1pt} } \partial _{\mu _{1} ...\mu _{p} }^{(p)} g^{a}  ,                      \]
that due to the arbitrariness of the functions $g^{a} (x)$, which parameterize the group $G_{M}^{g} $, gives:

\begin{equation} \label{eq11}
[L]{\kern 1pt} _{I} {\kern 1pt} {\kern 1pt} a{\kern 1pt} _{\, \, a}^{I{\kern 1pt} {\kern 1pt} \mu _{1} ...\mu _{p} } =S{\kern 1pt} _{a}^{\{ \mu _{1} ...\mu _{p} {\kern 1pt} \} } \quad {\rm if} \quad        0\le p\le m \, ,
\end{equation}

\begin{equation} \label{eq12}
S{\kern 1pt} _{a}^{\{ \mu _{1} ...\mu _{p} {\kern 1pt} \} } =0 \quad {\rm if} \quad  m<p\le m+n\, ,
\end{equation}
where braces mean symmetrization by the indices contained in them.

Identities (\ref{eq11}), (\ref{eq12}) are strong, i.e. they are fulfilled exclusively due to the $G_{M}^{g} $-invariance of the theory without assumption about the extremeness of the action.

On shell $[L]{\kern 1pt} _{I} =0$ relation (\ref{eq12}) holds for all $p$ (weak identities). For $p=0$, 1 and 2 it gives:

\begin{equation} \label{eq15} \partial _{\sigma } J{\kern 1pt} _{a}^{{\kern 1pt} \sigma } =0$, \;  $ S{\kern 1pt} _{a}^{\mu {\kern 1pt} } =J{\kern 1pt} _{a}^{\mu {\kern 1pt} } +\partial _{\sigma } J{\kern 1pt} _{a}^{\mu {\kern 1pt} \sigma } =0 $, \;  $ S{\kern 1pt} _{a}^{\{\mu \nu \}{\kern 1pt} } =J{\kern 1pt} _{a}^{\{\mu \nu\} {\kern 1pt} } +\partial _{\sigma } J{\kern 1pt} _{a}^{\{\mu \nu \} {\kern 1pt} \sigma } =0 .
 \end{equation}
First of them miens that \textit{gauge charges} \textit{$Q_{a} =\int _{V} J_{a}^{0} dV $} associated with the currents $J{\kern 1pt} _{a}^{{\kern 1pt} \sigma } $ are conserved $\partial _{0} Q_{a} =0$ (here $x^{0} $ is the time coordinate, and $dV$ is an element of 3-dimensional volume $V$). Provided
\begin{equation} \label{eq10} \partial _{\nu \sigma }^{2} J{\kern 1pt} _{a}^{\mu \nu  {\kern 1pt} \sigma }=0
\end{equation}
the rest of identities (\ref{eq15}) gives:
\begin{equation} \label{eq8}
J{\kern 1pt} _{a}^{\mu {\kern 1pt} } =-\partial _{\sigma } S{\kern 1pt} _{a}^{\mu {\kern 1pt} \sigma } $, \qquad  $ S{\kern 1pt} _{a}^{\mu {\kern 1pt} \sigma } =-S{\kern 1pt} _{a}^{\sigma {\kern 1pt} \mu } .
\end{equation}
So gauge charges in this case are quasilocal $ Q_{a} =\oint_{\partial V} S{\kern 1pt} _{a}^{i {\kern 1pt} 0 } ds_i $ and quantities $ S{\kern 1pt} _{a}^{\mu \nu {\kern 1pt} } =J{\kern 1pt} _{a}^{\mu \nu {\kern 1pt} } +\partial _{\sigma } J{\kern 1pt} _{a}^{\mu \nu {\kern 1pt} \sigma } $ acts as their \textit{superpotentials}. Here $ds_i $ is a vector of elementary area of the surface $\partial V$. Using the right to arbitrarily choose the sequence of the first two indices in $J{\kern 1pt} _{a}^{\mu \nu  {\kern 1pt} \sigma }$, index $\nu$ with which the condition (\ref{eq10}) is satisfied will be located in the penultimate place. \textit{Quasilocality condition} (\ref{eq10}) is performed, for example, in the case $J{\kern 1pt} _{a}^{\mu \{ \nu  {\kern 1pt} \sigma \} }=0$. Note that if condition (\ref{eq10}) is satisfied, we do not need $ J{\kern 1pt} _{a}^{\mu _{1} ...\mu _{p} {\kern 1pt}  }  $ with $ p > 3 $ to determine superpotentials $ S{\kern 1pt} _{a}^{\mu \nu {\kern 1pt} } $.

	The $G_{M}^{g}$-invariant theory with quasilocality condition (\ref{eq10}) we will call \textit{the gauge theory of the generalized gauge group} $G_{M}^{g}$.

We concretize now the obtained relations for $m=1$. In this case

\begin{equation} \label{eq16} \delta {\kern 1pt} q^{I} =a{\kern 1pt} _{\, a}^{I{\kern 1pt} {\kern 1pt} } g^{a} +b{\kern 1pt} _{\, \, a}^{I{\kern 1pt} {\kern 1pt} \mu } \partial _{\mu }^{} g^{a}  \end{equation}
(here the coefficients $a{\kern 1pt} _{\, \, a}^{I{\kern 1pt} {\kern 1pt} \mu } $ from formula (\ref{eq4}) are denoted as $b{\kern 1pt} _{\, \, a}^{I{\kern 1pt} {\kern 1pt} \mu } $):

\begin{equation} \label{eq17} \eqalign{ {J{\kern 1pt} _{a}^{\mu _{1} ...\mu _{p} {\kern 1pt} \sigma } =-\sum _{k=p-1}^{n-1}C_{k}^{p-1} {\kern 1pt} [L]{\kern 1pt} _{I}^{ \mu _{1} ...\mu _{p-1} \nu _{p} ...\nu _{k} {\kern 1pt} \sigma } \partial _{\nu _{p} ...\nu _{k} }^{(k-p+1)} b{\kern 1pt} _{\, \, a}^{I{\kern 1pt} {\kern 1pt} \mu _{p}  }  } \cr {\quad \quad \quad \quad \; \; \; -\sum _{k=p}^{n-1}C_{k}^{p} {\kern 1pt} [L]{\kern 1pt} _{I}^{\mu _{1} ...\mu _{p} \nu _{p+1} ...\nu _{k} {\kern 1pt} \sigma } \partial _{\nu _{p+1} ...\nu _{k} }^{(k-p)} a{\kern 1pt} _{\, \, a}^{I{\kern 1pt} {\kern 1pt} }  } } \end{equation}
in particular

\begin{equation} \label{eq18} J{\kern 1pt} _{a}^{\mu {\kern 1pt} {\kern 1pt} \sigma } =-\sum _{k=0}^{n-1}[L]{\kern 1pt} _{I}^{\nu _{1} ...\nu _{k} {\kern 1pt} \sigma } \partial _{\nu _{1} ...\nu _{k} }^{(k)} b{\kern 1pt} _{\, \, a}^{I{\kern 1pt} {\kern 1pt} \mu }  -\sum _{k=1}^{n-1}k{\kern 1pt} \, [L]{\kern 1pt} _{I}^{\mu {\kern 1pt} {\kern 1pt} \nu _{2} ...\nu _{k} {\kern 1pt} \sigma } \partial _{\nu _{2} ...\nu _{k} }^{(k-1)} a{\kern 1pt} _{\, a}^{I{\kern 1pt} {\kern 1pt} } \,,  \end{equation}

\begin{equation} \label{eq19} \eqalign{ {J{\kern 1pt} _{a}^{\mu {\kern 1pt} \rho {\kern 1pt} \sigma } =-\sum _{k=1}^{n-1}k{\kern 1pt} \, [L]{\kern 1pt} _{I}^{ \mu {\kern 1pt} {\kern 1pt} \nu _{2} ...\nu _{k} {\kern 1pt} \sigma } \partial _{\nu _{2} ...\nu _{k} }^{(k-1)} b{\kern 1pt} _{\, {\kern 1pt} a}^{I\, {\kern 1pt} {\kern 1pt} \rho  }} \cr {\quad \quad \quad \;\;  -\sum _{k=2}^{n-1}\frac{k(k-1)}{2} {\kern 1pt} \, [L]{\kern 1pt} _{I}^{\mu {\kern 1pt} \rho {\kern 1pt} \nu _{3} ...\nu _{k} {\kern 1pt} \sigma } \partial _{\nu _{3} ...\nu _{k} }^{(k-2)} a{\kern 1pt} _{\, a}^{I\, }  } }  \end{equation}
(expression (\ref{eq6}) for currents $J{\kern 1pt} _{a}^{\sigma } $ remains unchanged).

For $n=1$, formulas (\ref{eq6}), (\ref{eq18}) and (\ref{eq19})  give:

\begin{equation} \label{eq24} J{\kern 1pt} _{a}^{\sigma } =-\partial _{I}^{\sigma } L\, a{\kern 1pt} _{\, a}^{I{\kern 1pt} {\kern 1pt} } -L{\kern 1pt} h_{a}^{\sigma } ,                                                                                                \end{equation}

\begin{equation} \label{eq25} J{\kern 1pt} _{a}^{\mu {\kern 1pt} {\kern 1pt} \sigma } =S{\kern 1pt} _{a}^{\mu {\kern 1pt} {\kern 1pt} \sigma } =-\partial _{I}^{\sigma } L\, b{\kern 1pt} _{\, \, a}^{I{\kern 1pt} {\kern 1pt} \mu } $, \qquad  $ J{\kern 1pt} _{a}^{\mu {\kern 1pt} \rho {\kern 1pt} \sigma } =0
\end{equation}
and for $n=2$ give:

\begin{equation} \label{eq13}
J{\kern 1pt} _{a}^{\sigma } =-\, [L]{\kern 1pt} _{I}^{{\kern 1pt} \sigma } a{\kern 1pt} _{\, \, a}^{I{\kern 1pt} } -\partial {\kern 1pt} _{I}^{\nu \, {\kern 1pt} \sigma } L\, \, \partial _{\nu }^{} \, a{\kern 1pt} _{\, \, a}^{I{\kern 1pt} } -L\, {\kern 1pt} h_{a}^{\sigma } ,                                                                               \end{equation}
\begin{equation} \label{eq14}
J{\kern 1pt} _{a}^{\mu {\kern 1pt} {\kern 1pt} \sigma } =-\, [L]{\kern 1pt} _{I}^{{\kern 1pt} \sigma } b{\kern 1pt} _{\, \, a}^{I{\kern 1pt} {\kern 1pt} \mu } -\partial {\kern 1pt} _{I}^{\nu \, {\kern 1pt} \sigma } L\, \, \partial _{\nu }^{} \, b{\kern 1pt} _{\, \, a}^{I{\kern 1pt} {\kern 1pt} \mu } -\partial {\kern 1pt} _{I}^{\mu {\kern 1pt} {\kern 1pt} {\kern 1pt} \sigma } L\, \, a{\kern 1pt} _{\, a}^{I{\kern 1pt} {\kern 1pt} } ,
\end{equation}
\begin{equation} \label{eq22}
J{\kern 1pt} _{a}^{\mu {\kern 1pt} \rho {\kern 1pt} \sigma } =-{\kern 1pt} \, \partial {\kern 1pt} _{I}^{ \mu {\kern 1pt} {\kern 1pt} {\kern 1pt} \sigma } L\, \, b{\kern 1pt} _{\, {\kern 1pt} a}^{I\, {\kern 1pt} {\kern 1pt} \rho  } .                                                                                                     \end{equation}

Provided that the quasilocality condition (\ref{eq10}) is met, identities (\ref{eq11}) in our case are written as follows (strong identities):

\begin{equation} \label{eq20} [L]{\kern 1pt} _{I} {\kern 1pt} {\kern 1pt} a{\kern 1pt} _{\, \, a}^{I{\kern 1pt} {\kern 1pt} } =\partial _{\sigma } J{\kern 1pt} _{a}^{{\kern 1pt} \sigma } \,, \end{equation}

\begin{equation} \label{eq21} [L]{\kern 1pt} _{I} {\kern 1pt} {\kern 1pt} b{\kern 1pt} _{\, \, a}^{I{\kern 1pt} {\kern 1pt} \mu } =J{\kern 1pt} _{a}^{\mu {\kern 1pt} } +\partial _{\sigma } S{\kern 1pt} _{a}^{\mu {\kern 1pt} \sigma } \,. \end{equation}

	In gauge theories, the deformation coefficients $ h_{\mu}^{a}$ of generalized gauge deformed groups $G_{M}^{g}$ act as potentials (Appendix), gauge transformations for them (\ref{eq102}), (\ref{eq115}) have the form:
\[
\delta h_{\mu}^{a}=a_{\mu b}^{a}\,g^b-\partial_\mu g^a
\]
and transformations all other fields of theory not contain derivatives of group parameters. So nonzero coefficients $ b_{\,\,a}^{I \sigma } $ in this case are only $ b_{\mu \, a}^{b \, \sigma }=-\delta_\mu ^\sigma \delta_a ^b $  (index $I$ now is multiindex ${}_\mu ^b$). Thereby identity (\ref{eq21}) gives:
\begin{equation} \label{eq23}
-[L]{\kern 1pt} _{a}^ \mu =J{\kern 1pt} _{a}^{\mu {\kern 1pt} } +\partial _{\sigma } S{\kern 1pt} _{a}^{\mu {\kern 1pt} \sigma }.
\end{equation}

So the following is true.

\textbf{Theorem}. \textit{In the gauge theory of the generalized gauge group $G_{M}^{g}$, the gauge charges $Q_{a} $ are quasilocal, i.e. their currents }$J{\kern 1pt} _{a}^{\mu {\kern 1pt} } $\textit{ have superpotentials $S{\kern 1pt} _{a}^{\mu \nu {\kern 1pt} } =J{\kern 1pt} _{a}^{\mu \nu {\kern 1pt} } +\partial _{\sigma } J{\kern 1pt} _{a}^{\mu \nu {\kern 1pt} \sigma } =-S{\kern 1pt} _{a}^{\nu \mu {\kern 1pt} } $:  $J{\kern 1pt} _{a}^{\mu {\kern 1pt} } =-\partial _{\sigma } S{\kern 1pt} _{a}^{\mu {\kern 1pt} \sigma } $. Moreover, expression of currents through superpotentials is equivalent to equations of the gauge field.}   \textit{}

This generalizes the corresponding theorem, which was proved in \cite{Sam Gologr}, to the Lagrangians with higher derivatives of fields.

From this theorem, in particular, it follows that the gauge field equations in the arbitrary gauge theory of the generalized gauge group $G_{M}^{g}$ can be written in the form of dynamic Maxwell equations (or Young-Mills equations).

\section{Laws of motion of uncharged matter}

Under the theory of gravity we will understand the gauge theory of the translations group, which is the group of Riemannian translations $T_{M}^{g} $ (Appendix A.3). The gravitational field potentials are identified with deformation coefficients $h_{\mu }^{m} $, which form inverse matrix to matrix $h_{m}^{\mu } $ of coefficients of an orthonormal frame (tetrad) $X_{m} =h_{m}^{\mu } \partial _{\mu } $ (indexed by indices $m,n,p,s$) in coordinate frame $\partial _{\mu } $. Replacement of coordinate indices to frame indices and vice versa will be performed using matrices $h_{\mu }^{m} $ and $h_{m}^{\mu } $.

We assume that in addition to the gravitational field in space-time $M$ there are also matter fields $\psi ^{\xi } $ (these are all fields except the gravitational one). The field index is $\xi $. All matter fields are set in the local frame, and therefore are $T_{M}^{g} $-scalars.

The Lagrangian of the gravitational system is defined as the sum

\[L=\sqrt{g} L_{G} (h,\partial h,...,\partial ^{(n)} h)+\sqrt{g} L_{\psi } (h,\partial h,\psi ,\partial \psi ),          \]
where $g=\left|g_{\mu \nu } \right|$, $g_{\mu \nu } =h_{\mu }^{m} h_{\nu }^{n} \eta _{mn} $. In our case, the field variables $q^{I} $ split into system $\{ \, h_{\mu }^{m} ,\psi ^{\xi } \} $ which is equivalent to splitting the field index $I$ into the multiindex $\{ \, _{\mu }^{m} ,\, ^{\xi } \} $. With the minimal method of inclusion of the gravitational interaction, $L_{\psi } $ does not depend on $\partial h$.

In this article, we will not specify either the Lagrangian of gravitational field $\sqrt{g} L_{G} $ or the Lagrangian of matter $\sqrt{g} L_{\psi } $, assuming only gauge translational invariance ($T_{M}^{g} $-invariance) of both $L_{G} $ and $L_{\psi } $, which provides equality to zero of integral variations of both components of the Lagrangian $L$ at $T_{M}^{g} $-transformations $\delta '(\sqrt{g} L_{G} )=0$, $\delta '(\sqrt{g} L_{\psi } )=0$.

Infinitesimal transformations of coordinates and field variables under the action of the group $T_{M}^{g} $ are follows (\ref{eq115}):

\begin{equation} \label{eq26} \delta x^{\mu } =h_{n}^{\mu } t^{n} $, \quad  $\delta {\kern 1pt} h_{\mu }^{m} =-F_{\mu {\kern 1pt} n}^{m} {\kern 1pt} t^{n} -\partial _{\mu } t^{m} $,  \quad  $\delta \psi ^{\xi } =-\partial _{n} {\kern 1pt} \psi ^{\xi } t^{n} .                                              \end{equation}
Thus in our case

\begin{equation} \label{eq27} a_{\mu {\kern 1pt} n}^{m} =-F_{\mu {\kern 1pt} n}^{m} $\,,   \quad   $a^{\xi } _{n} =-\partial _{n} {\kern 1pt} \psi ^{\xi } $\,,  \quad  $b_{\mu \, n}^{m\nu } =-\delta _{n}^{m} \delta _{\mu }^{\nu }\, ,
\end{equation}
and all other coefficients $b_{\; \, m}^{I\, \, \mu } =0$. Here $\partial _{n} :=h_{n}^{\mu } \partial _{\mu } $.

We introduce now the notation for variational derivatives. Let be

\begin{equation} \label{eq28} [\sqrt{g} L_{G} ]_{m}^{\mu } =:\sqrt{g} G_{m}^{\mu } ,                                                                                                            \end{equation}
where $G_{m}^{\mu } $ is a generalized Einstein tensor, which is calculated by formula (\ref{eq2}) with $q^{I} \to h_{\mu }^{m} $:

\begin{equation} \label{eq29} \eqalign{  {\sqrt{g} G_{m}^{\mu } =\sum _{k=0}^{n}(-1)^{k} \partial _{\sigma _{1} ...\sigma _{k} }^{(k)} \partial {\kern 1pt} _{m}^{\mu ,\, \sigma _{1} ...\sigma _{k} } (\sqrt{g} L_{G} ) } \cr {\quad \quad \quad  = \partial {\kern 1pt} _{m}^{\mu } (\sqrt{g} L_{G} )-\partial _{\sigma }^{} (\sqrt{g} {\kern 1pt} \partial {\kern 1pt} _{m}^{\mu ,\sigma } L_{G} )+\partial _{\sigma {\kern 1pt} \rho }^{(2)} (\sqrt{g} {\kern 1pt} \partial {\kern 1pt} _{m}^{\mu ,\sigma {\kern 1pt} \rho } L_{G} )-...} } \end{equation}
Here it is accepted that at $q^{I} =h_{\mu }^{m} $: $\partial _{I} =\partial {\kern 1pt} _{m}^{\mu } $, $\partial {\kern 1pt} _{I}^{\nu _{1} ...\nu _{l} } =\partial {\kern 1pt} _{m}^{\mu ,\nu _{1} ...\nu _{l} } $, $[L]{\kern 1pt} _{I} =[L]{\kern 1pt} _{m}^{\mu } $, $[L]{\kern 1pt} _{I}^{\nu _{1} ...\nu _{l} } =[L]{\kern 1pt} _{m}^{\mu ,\nu _{1} ...\nu _{l} } $. In addition

\begin{equation} \label{eq30} [\sqrt{g} L_{\psi } ]{\kern 1pt} _{\xi } =:\sqrt{g} {\kern 1pt} G_{\xi }\, ,                                                                                                              \end{equation}
where ${\kern 1pt} G_{\xi }^{} =f_{\xi } -\nabla _{\sigma } p_{\xi }^{\sigma } $, $f_{\xi } :=\partial _{\xi }^{} L_{\psi } $, $p_{\xi }^{\sigma } :=\partial _{\xi }^{\sigma } L_{\psi } $ and $\nabla _{\sigma } $ is a covariant derivative in Riemannian space with metric $g_{\mu \nu } $: $\nabla _{\sigma } p_{\xi }^{\sigma } ={\frac{1}{\sqrt{g} }} \partial _{\sigma } (\sqrt{g} {\kern 1pt} p{\kern 1pt} _{\xi }^{\sigma } )$. Next, we introduce the notation

\begin{equation} \label{eq31} [\sqrt{g} L_{\psi } ]_{m}^{\mu } =:-\sqrt{g} {\kern 1pt} \tau _{m}^{\mu }\, ,                                                                                                           \end{equation}
so

\begin{equation} \label{eq32} \tau _{m}^{\mu } =\sigma _{m}^{\mu } +\nabla _{\sigma } \beta _{m}^{\mu \, \sigma } $, \quad $\sqrt{g} {\kern 1pt} \sigma _{m}^{\mu } :=-\partial _{m}^{\mu } (\sqrt{g} L_{\psi } )$, \quad $\beta _{{\kern 1pt} \, m}^{\, \mu {\kern 1pt} {\kern 1pt} \sigma } :=\, \partial _{m}^{\mu ,\sigma } L_{\psi }\, ,
\end{equation}
where $\nabla _{\sigma } \beta {\kern 1pt} _{m}^{\mu \sigma } ={\frac{1}{\sqrt{g} }} \partial _{\sigma } (\sqrt{g} \beta {\kern 1pt} _{m}^{\mu \sigma } )$. This follows from the fact that the tensor density $S_{\psi {\kern 1pt} \, n}^{\, \; \, \nu {\kern 1pt} {\kern 1pt} \sigma } =\, \sqrt{g} \beta _{\, n}^{{\kern 1pt} \nu \, \sigma } $ is a translational superpotential of the Lagrangian $\sqrt{g} L_{\psi } $ and therefore is an antisymmetric quantity (with respect to upper indices) due to the $T_{M}^{g} $-invariance of $L_{\psi } $.

It is shown below that the mixed coordinate-frame tensor $\tau _{m}^{\mu } $ can be interpreted as the energy-momentum tensor of matter. So in the accepted designations

\[[L]_{m}^{\mu } =\sqrt{g} (G_{m}^{\mu } -\tau _{m}^{\mu } ), \]
and equations of the gravitational field $[L]_{m}^{\mu } =0$ in any frame theory of gravity (where $h_{\mu }^{m} $ are the gravitational potentials) can be written in the form of Einstein equation $G_{m}^{\mu } =\tau _{m}^{\mu } $.

Components of currents $J_{m}^{\mu } =J_{G{\kern 1pt} {\kern 1pt} m}^{\; \; \, \, \mu } +J_{\psi {\kern 1pt} m}^{\; \; \mu } $ and superpotentials $S_{m}^{\mu \nu } =S_{G{\kern 1pt} {\kern 1pt} m}^{\; \; \mu \nu } +S_{\psi {\kern 1pt} m}^{\; {\kern 1pt} \mu \nu } $ are associated with the corresponding components $\sqrt{g} L_{G} $ and $\sqrt{g} L_{\psi } $ of the Lagrangian $L$.

To find the components associated with the gravitational Lagrangian, we specify expressions (\ref{eq6}), (\ref{eq18}), (\ref{eq19}) for $L=\sqrt{g} L_{G} $, taking into account formulas (\ref{eq27}):

\begin{equation} \label{eq33} J_{G\, n}^{\; \; \, \sigma } =\sum _{k=0}^{n-1}[\sqrt{g} L_{G} ]{\kern 1pt} _{m}^{\lambda ,\nu _{1} ...\nu _{k} {\kern 1pt} \sigma } \partial _{\nu _{1} ...\nu _{k} }^{(k)} F{\kern 1pt} _{\lambda \, n}^{\, m{\kern 1pt} {\kern 1pt} }  -\sqrt{g} L_{G} {\kern 1pt} h_{n}^{\sigma }\,,  \end{equation}

\begin{equation} \label{eq34} J_{G\, n}^{\; \; \mu {\kern 1pt} {\kern 1pt} \sigma } =[\sqrt{g} L_{G} ]{\kern 1pt} _{n}^{\mu ,\, \sigma } +\sum _{k=1}^{n-1}k{\kern 1pt} \, [\sqrt{g} L_{G} ]{\kern 1pt} _{m}^{\lambda ,\mu {\kern 1pt} {\kern 1pt} \nu _{2} ...\nu _{k} {\kern 1pt} \sigma } \partial _{\nu _{2} ...\nu _{k} }^{(k-1)} F{\kern 1pt} _{\lambda \, n}^{\, m{\kern 1pt} {\kern 1pt} } \,,  \end{equation}

\begin{equation} \label{eq35} \; J_{G\, n}^{\; \; \mu {\kern 1pt} \rho {\kern 1pt} \sigma } =\, [\sqrt{g} L_{G} ]{\kern 1pt} _{\, \, n}^{ \mu {\kern 1pt} ,\rho {\kern 1pt} \sigma } +\sum _{k=2}^{n-1}\frac{k(k-1)}{2} {\kern 1pt} \, [\sqrt{g} L_{G} ]{\kern 1pt} _{m}^{\lambda ,\mu {\kern 1pt} {\kern 1pt} \rho \, \nu _{3} ...\nu _{k} {\kern 1pt} \sigma } \partial _{\nu _{3} ...\nu _{k} }^{(k-2)} F{\kern 1pt} _{\lambda \, n}^{\, m{\kern 1pt} {\kern 1pt} } \, .
\end{equation}
For $n = 1$ we have:
\[J_{G\, n}^{\; \; \, \sigma } =\sqrt{g} {\kern 1pt} {\kern 1pt} (\partial {\kern 1pt} _{m}^{\lambda ,\, \sigma } L_{G} \, F{\kern 1pt} _{\lambda \, n}^{\, m{\kern 1pt} {\kern 1pt} } -L_{G} {\kern 1pt} h_{n}^{\sigma } )\,,\]

\[J_{G\, n}^{\; \; \mu {\kern 1pt} {\kern 1pt} \sigma } =\sqrt{g} {\kern 1pt} \partial {\kern 1pt} _{n}^{\mu ,\, \sigma } L_{G}\,, \quad J_{G\, n}^{\; \; \mu {\kern 1pt} \rho {\kern 1pt} \sigma } =\, 0\,.                                                                                                                      \]
and for $n=2$:
\[J_{G\, n}^{\; \; \, \sigma } =[\sqrt{g} L_{G} ]{\kern 1pt} _{m}^{\lambda ,{\kern 1pt} {\kern 1pt} \sigma } F{\kern 1pt} _{\lambda \, n}^{\, m{\kern 1pt} {\kern 1pt} } +\sqrt{g}\, \partial{\kern 1pt} _{m}^{\lambda ,\nu {\kern 1pt} \sigma } L_{G} \, \partial _{\nu }^{} F{\kern 1pt} _{\lambda \, n}^{\, m{\kern 1pt} {\kern 1pt} } -\sqrt{g} L_{G} {\kern 1pt} h_{n}^{\sigma }\,, \]

\[J_{G\, n}^{\; \; \mu {\kern 1pt} {\kern 1pt} \sigma } =[\sqrt{g} L_{G} ]{\kern 1pt} _{n}^{\mu ,\, \sigma } +\sqrt{g}\, \partial{\kern 1pt} _{m}^{\lambda ,\mu {\kern 1pt} \sigma } L_{G} \, F{\kern 1pt} _{\lambda \, n}^{\, m{\kern 1pt} {\kern 1pt} },\quad                                                                            \; J_{G\, n}^{\; \; \mu {\kern 1pt} \rho {\kern 1pt} \sigma } =\, \sqrt{g} \, \partial _{\, \, n}^{\, \rho,\mu  {\kern 1pt} \sigma } L_{G}\, .                                                                                                 \]

For theories with $n = 1$, the condition of quasilocality (\ref{eq10}) due to $J_{G\, n}^{\; \; \mu {\kern 1pt} \rho {\kern 1pt} \sigma } =\, 0$ is obviously satisfied. This condition also holds for all $f(\Re)$-theories ($n=2$), where $\Re$ is the curvature tensor $R_{\: \rho \, \sigma \mu} ^ n$. Indeed, in this case $ J_{G\, n}^{\; \; \mu {\kern 1pt} \rho {\kern 1pt} \sigma }=D_{\quad \,\, n} ^{\rho \, \sigma \mu}+D_n ^{\: \, \rho \, \sigma \mu}-D_n ^{\: \, \sigma \rho \, \mu} \,$, where $D_n ^{\: \, \rho \, \sigma \mu}:=4 \, \partial f(\Re)/\partial R_{\: \rho \, \sigma \mu} ^ n$, so $J_{G\, n}^{\; \; \mu {\kern 1pt} \{ \rho {\kern 1pt} \sigma \} } =\, 0$.

We enter the notation:

\[{\kern 1pt} \sqrt{g} {\kern 1pt} t_{m}^{\mu } :=J_{G{\kern 1pt} m}^{\; \, \, \, \mu } ,  \qquad  \sqrt{g} B{\kern 1pt} _{m}^{\mu \nu } :=S_{G{\kern 1pt} m}^{\; \, \mu \nu } =J_{G{\kern 1pt} m}^{\; \, \mu \nu } +\partial _{\sigma } J_{G{\kern 1pt} a}^{\, \, \mu \nu {\kern 1pt} \sigma } .                 \]

Both Noether's currents $J_{m}^{\mu } $ and corresponding superpotentials $S{\kern 1pt} _{m}^{\mu \sigma } $ in terms of coordinate transformations ($T_{M}^{g} $-transformations) are tensor densities, which when divided by $\sqrt{g} $ become tensors (of appropriate rank), which we will call the \textit{tensor currents} and the \textit{tensor superpotentials}, respectively. By the frame index in relation to the global Lorentz transformations, they are vectors, so by all indices in relation to both $T_{M}^{g} $-transformations and global Lorentz transformations, they are mixed tensors. Thus, the tensor current ${\kern 1pt} t_{m}^{\mu } $ associated with the gauge translational invariance of Lagrangian $L_{G} $ is mixed \textit{energy-momentum tensor of the gravitational field} (in the theory of gravity described by the Lagrangian $L_{G} $), and $B{\kern 1pt} _{m}^{\mu \sigma } $ is its tensor superpotential which we will call \textit{tensor of induction of the gravitational field}. According to its definition and formulas (\ref{eq34}), (\ref{eq35}), this tensor is determined by a specific type of Lagrangian $ L_G $ of gravitational theory. So for $f(\Re)$-theories, it is given by the expression:
\[B_{m}^{\mu \nu }=D_s ^{\: \, n \mu \nu}\omega_{\: mn}^s-G_m^{\:\,[\mu \nu]}+G_{\:\,\,m}^{[\mu \:\: \nu]}+G_{\:\,\:\,\:\,m}^{[\mu \nu]}
\]
where $G_m^{\,\: \, n \nu}:=D_m ^{\: \, \, \, s \sigma \nu}\omega_{\: \sigma s}^n-D_s ^{\: \, n \sigma \nu}\omega_{\: \sigma m}^s-\nabla_\sigma D_m ^{\: \, \,\, n \nu \sigma }$ are variational derivations of $L_G$ with respect to spin connection $\omega_{\: \nu n}^m$, hence it is determined by the quantity $D_m ^ {\: \, \, n\mu\nu}$, which is different for different $f(\Re)$-theories. For example, in the case Einstein-Gauss-Bonnet theory $f(\Re)=R+\alpha(R^2-4 R_{\mu\nu}R^{\mu\nu}+R_{\mu\nu\rho\sigma}R^{\mu\nu\rho\sigma})$, therefore
\[ D_{mn}^{\quad \: \mu\nu}=h_m^{[\mu} h_n^{\nu]}+2\alpha (R h_m^{[\mu} h_n^{\nu]}-4 h_{[m}^{[\mu} h_{|k|}^{\,\nu]}R_{n]}^{\,k}+R_{mn}^{\quad \: \mu\nu}).
\]
Square brackets here mean antisymmetrization by the indexes contained in them, vertical bars highlight indices that are skipped during antisymmetrization. In GR with Hilbert’s Lagrangian $f(\Re)=R$, $D_{mn}^{\quad \: \mu\nu}=h_m^{[\mu} h_n^{\nu]}$, $G_m^{\,\: \, n \nu}=0$, so $B_{m}^{\mu \nu }=\omega _{\,\,m\, }^{\,\mu \:\,\nu }$. In the theory of gravity with truncated Hilbert’s Lagrangian (M{\o}ller’s Lagrangian) $L_G=-\frac{1}{2}R-\nabla_\sigma R^\sigma$, where $R_n:=F_{s n}^s$, we have
$B_{m}^{\mu \nu }=\omega _{\,\,m\, }^{\,\mu \:\,\nu }-h_{m}^{\mu }\,{{R}^{\,\nu }}+h_{m}^{\nu }\,{{R}^{\,\mu }}$
\cite{Sam ClQGr}. This expression is used in teleparallel equivalent of GR.

For the Lagrangian $\sqrt{g} L_{\psi } $, the translation current and its superpotential are given by the formulas:

\[J{\kern 1pt} _{\psi {\kern 1pt} n}^{\, \; \, \, \nu } =\sqrt{g} (\beta _{m}^{{\kern 1pt} \mu \, \nu } F{\kern 1pt} _{\mu {\kern 1pt} n}^{m{\kern 1pt} {\kern 1pt} } +\, p_{\xi }^{\nu } \partial _{n} \psi ^{\xi } {\kern 1pt} {\kern 1pt} -L_{\psi } {\kern 1pt} h_{n}^{\nu } ),            \qquad S_{\psi {\kern 1pt} \, n}^{\, \; \, \nu {\kern 1pt} {\kern 1pt} \sigma } =\, \sqrt{g} \beta _{\, n}^{{\kern 1pt} \nu \, \sigma } ,\]
which specify for this case formulas (\ref{eq24}) and (\ref{eq25}), respectively.

Consider the identity (\ref{eq21}) for both components of the Lagrangian $L$. For $\sqrt{g} L_{G} $ we have:

\begin{equation} \label{eq36} -G_{m}^{\mu } =t{\kern 1pt} _{m}^{\mu } +\nabla _{\sigma } B{\kern 1pt} _{m}^{\mu \sigma } ,                                                                                                          \end{equation}
and for $\sqrt{g} L_{\psi } $:

\begin{equation} \label{eq37} \, \tau _{n}^{\nu } =J{\kern 1pt} _{\psi {\kern 1pt} n}^{\, \; \, \, \nu } /\sqrt{g} +\, \nabla _{\sigma } \beta _{\, n}^{{\kern 1pt} \nu {\kern 1pt} \sigma } .                                                                                                  \end{equation}
Comparison of the latter identity with expression (\ref{eq32}) gives a convenient way to calculate the translational Noether's current of the Lagrangian of matter:

\begin{equation} \label{eq38} J{\kern 1pt} _{\psi {\kern 1pt} n}^{\, \; \, \, \nu } =\sqrt{g} \, \sigma _{n}^{\nu } =-\partial _{n}^{\,\nu } (\sqrt{g} L_{\psi } )\,.                                                                                             \end{equation}
Therefore, the mixed coordinate-frame tensor $\sigma _{n}^{\nu } $ is the energy-momentum tensor of matter fields. The tensor $\tau _{n}^{\nu } $ differs from $\sigma _{n}^{\nu } $ by the covariant divergence of the antisymmetric tensor $\beta _{\, n}^{{\kern 1pt} \nu {\kern 1pt} \sigma } $, so it can also be interpreted as the energy-momentum tensor of matter. At the minimal method of including the gravitational interaction $\beta _{\, n}^{{\kern 1pt} \nu {\kern 1pt} \sigma } =0$, therefore both tensors $\tau _{n}^{\nu } $ and $\sigma _{n}^{\nu } $ coincide.

 From identity (\ref{eq36}) it follows:

\begin{equation} \label{eq39} -[L]_{m}^{\mu } /\sqrt{g} =-(G_{m}^{\mu } -\tau _{m}^{\mu } )=T{\kern 1pt} _{m}^{\mu } +\nabla _{\sigma } B{\kern 1pt} _{m}^{\mu \sigma } ,                                                                        \end{equation}
where $T{\kern 1pt} _{m}^{\mu } \, :=t_{m}^{\mu } +\tau _{m}^{\mu } $ is the total energy-momentum tensor of the gravitational system. The obtained identity makes it possible to write equations of the gravitational field $[L]_{m}^{\mu } =0$ in the superpotential form:

\begin{equation} \label{eq40} \nabla _{\sigma } B{\kern 1pt} _{m}^{\mu \sigma } =-T{\kern 1pt} _{m}^{\mu } \,,                                                                                                                   \end{equation}
similar to the form of Maxwell's dynamic equations (and also the form of equations for other gauge fields of internal symmetry). As we can see, this possibility is provided exclusively by the gauge translational invariance of the theory of gravity, and not by a specific type of gravitational Lagrangian $ L_{G} $.

So according to Theorem (Section 2) equations of the gravitational field $[L]_{m}^{\mu } =0$ in any gauge theory of the generalized gauge group $T_{M}^{g}$ can be written both in the form of Einstein equation $G_{m}^{\mu } =\tau _{m}^{\mu }$ and in the superpotential form $\nabla _{\sigma } B{\kern 1pt} _{m}^{\mu \sigma } =-T{\kern 1pt} _{m}^{\mu } $, which is an expression of the total mixed coordinate-frame energy-momentum tensor of the gravitational system $T{\kern 1pt} _{m}^{\mu } $ through the covariant divergence of the tensor of induction of the gravitational field $B{\kern 1pt} _{m}^{\mu \sigma } $, which is its tensor superpotential. The tensor of induction of the gravitational field $B{\kern 1pt} _{m}^{\mu \sigma } $ is determined by the gravitational Lagrangian $ L_{G} $ and acts as \textit{power characteristic of the gravitational field} in the theory of gravity based on $ L_{G} $. It is born by the energy-momentum of the gravitational system and characterizes the gravitational field better than curvature or torsion. Being a tensor superpotential of energy-momentum, tensor $ B_m ^ {\mu\nu} $ allows us to find the energy-momentum vector of a gravitational system closed in a 3-dimensional volume $V$ as an integral over the 2-dimensional boundary $\partial V$ of a given volume $ P_{m} =\oint_{\partial V} B{\kern 1pt} _{m}^{i {\kern 1pt} 0 } \sqrt{g} \,ds_i $, which shows the holographic nature of gauge theories of gravity.

We turn to the consideration of identity (\ref{eq20}). For $\sqrt{g} L_{G} $ it is reduced to

\begin{equation} \label{eq41} -G{\kern 1pt} _{m}^{\mu } {\kern 1pt} {\kern 1pt} F{\kern 1pt} _{\mu \, n}^{m{\kern 1pt} {\kern 1pt} } =\nabla _{\sigma } t{\kern 1pt} _{n}^{\sigma } ,                                                                                                             \end{equation}
or taking into account identity (\ref{eq36}) to

\begin{equation} \label{eq42} (t{\kern 1pt} _{m}^{\mu } +\nabla _{\sigma } B{\kern 1pt} _{m}^{\mu \sigma } ){\kern 1pt} {\kern 1pt} F{\kern 1pt} _{\mu \, n}^{m{\kern 1pt} {\kern 1pt} } =\nabla _{\sigma } t{\kern 1pt} _{n}^{\sigma } ,                                                                                               \end{equation}
and for $\sqrt{g} L_{\psi } $ to

\begin{equation} \label{eq43} \tau _{m}^{\mu } {\kern 1pt} F{\kern 1pt} _{\mu \, n}^{m{\kern 1pt} {\kern 1pt} } -G_{\xi }^{} \partial _{n} \psi ^{\xi } =\nabla _{\sigma } \tau {\kern 1pt} _{n}^{\sigma } .                                                                                                \end{equation}

Adding (\ref{eq42}) and (\ref{eq43}), we obtain:

\begin{equation} \label{eq44} (T{\kern 1pt} _{m}^{\mu } +\nabla _{\sigma } B{\kern 1pt} _{m}^{\mu \sigma } ){\kern 1pt} {\kern 1pt} F{\kern 1pt} _{\mu \, n}^{m{\kern 1pt} {\kern 1pt} } -G_{\xi }^{} \partial _{n} \psi ^{\xi } =\nabla _{\sigma } T{\kern 1pt} _{n}^{\sigma } ,                                                                           \end{equation}

The identities discussed above are strong. Consider now their weak (weakened) version.

From the equations of the gravitational field, due to the antisymmetry of the tensor of induction of the gravitational field $B{\kern 1pt} _{m}^{\mu \sigma } $, follows the law of conservation of the total energy-momentum of the gravitational system:

\begin{equation} \label{eq45} -\nabla _{\mu } \nabla _{\sigma } B{\kern 1pt} _{m}^{\mu \sigma } =\nabla _{\mu } T{\kern 1pt} _{m}^{\mu } =\nabla _{\mu } t{\kern 1pt} _{m}^{\mu } +\nabla _{\mu } \tau {\kern 1pt} _{m}^{\mu } =0\,.
\end{equation}

On the gravitational extremal $G_{m}^{\mu } =\tau _{m}^{\mu } $, as well as $ -\nabla _{\mu } t{\kern 1pt} _{m}^{\mu } =\nabla _{\mu } \tau {\kern 1pt} _{m}^{\mu } $, so from identity (\ref{eq41}) we obtain the identity (weak)

\begin{equation} \label{eq46} \nabla _{\sigma } \tau {\kern 1pt} _{n}^{\sigma } =\tau _{m}^{\mu } F_{\mu \, n}^{m}\,,  \end{equation}
which is \textit{the equation of transfer of energy-momentum of matter, }or\textit{ the law of motion of matter}. Thus, the law of motion of matter (\ref{eq46}) is an identity which follows from equations of the gravitational field due to the gauge translational invariance. From (\ref{eq46}) follows that tensor $F{\kern 1pt} _{\mu \, n}^{m{\kern 1pt} {\kern 1pt} } $ describes \textit{strength of gravitational field} and $\tau _{m}^{\mu } {\kern 1pt} $ its \textit{gravitational charge}. Equality $F{\kern 1pt} _{\mu \, n}^{m{\kern 1pt} {\kern 1pt} }=0$ is a sign of flat space.

Consider the law of motion of matter (\ref{eq46}) in the case of dust matter, when its energy-momentum tensor is written as $\tau _{m}^{\mu } {\kern 1pt} =\pi ^{\mu } u_{m} $, where $u^{\mu} =dx^{\mu} /ds$ is the 4-velosity of particles and $\pi ^{\mu } =\mu {\kern 1pt} cu^{\mu } $ is the density of their 4-momentum ($\mu $ is the density of mass). Substituting this expression in (\ref{eq46}), we obtain:

\begin{equation} \label{eq47} \pi ^{\mu } \partial _{\mu } u_{m} =\pi ^{\mu } u_{n} F_{\mu m}^{n} -\nabla _{\mu } \pi ^{\mu } u_{m} .                                                                                      \end{equation}
When the mass of matter is conserved $\nabla _{\mu } \pi ^{\mu } =0$, the last term in (\ref{eq47}), which has the meaning of the density of reactive force, disappears. So the particles in this case move according to the law

\begin{equation} \label{eq48} u^{\mu } (\partial _{\mu } u_{m} -u_{n} F_{\mu m}^{n} )=0\,,                                                                                                      \end{equation}
which is the geodesic equation of Riemannian space with metric $g_{\mu \nu } =\eta _{mn} h_{\mu }^{m} h_{\nu }^{n}$, written in terms of 4-velocity $u_{m} $ and anholonomic coefficients $F_{\mu \nu }^{n} $.

On the gravitational extremal, identity (\ref{eq44}) is reduced to equations:

\begin{equation} \label{eq49} G_{\xi }^{} \partial _{n} {\kern 1pt} \psi ^{\xi } =0\,.                                                                                                                       \end{equation}
If the condition

\begin{equation} \label{eq50} rank{\kern 1pt} {\kern 1pt} (\partial _{n} \, \psi ^{\xi } )=f \end{equation}
is satisfied, where $f$ is the dimension of the field representation ($\xi =1,...f$), from (\ref{eq49}) follows  $G_{\xi }^{} =0$,  or ${\kern 1pt} \nabla _{\sigma } p_{\xi }^{\sigma } =f_{\xi } $. Therefore, if condition (\ref{eq50}) is satisfied, equations of matter fields follow from equations of the gravitational field. Condition (\ref{eq50}) can be satisfied only in the case $f\le 4$, in particular for scalar fields.

On the other hand, when equations $G_{\xi }^{} =0$ are fulfilled (on the extremal of matter fields) identity (\ref{eq43}) also gives the law of motion of matter (\ref{eq46}), and from identity (\ref{eq44}) it follows that with the additional assumption of the conservation of the total energy-momentum (\ref{eq45}), equations

\begin{equation} \label{eq51} (\nabla _{\nu } B_{n}^{\mu \nu } +T_{n}^{\mu } )F_{\mu m}^{n} =0 \end{equation}
are fulfilled. There are only 4 such equations, and therefore not all equations of the gravitational field follow from them, but only a certain part of them.

So the following is true.

 \textbf{Proposition 1.} \textit{In any gauge theory of the generalized gauge group} $T_{M}^{g}$:

\textit{ 1) when fulfilling equations of the gravitational field $\nabla _{\sigma } B{\kern 1pt} _{m}^{\mu \sigma } =-T{\kern 1pt} _{m}^{\mu } $ (on the gravitational extremal):}

\textit{ a) the law of motion of matter }$\nabla _{\sigma } \tau {\kern 1pt} _{n}^{\sigma } =\tau _{m}^{\mu } F_{\mu \, n}^{m} $ \textit{is fulfilled, according to which the dust particles move along geodesics of Riemannian space with metric $g_{\mu \nu } =\eta _{mn} h_{\mu }^{m} h_{\nu }^{n}$;}

\textit{ b) equations }$G_{\xi }^{} \partial _{n} {\kern 1pt} \psi ^{\xi } =0$\textit{ are fulfilled, which are equivalent to equations of matter fields }$G_{\xi }^{} =0$\textit{ if the condition }$rank{\kern 1pt} {\kern 1pt} (\partial _{n} {\kern 1pt} \psi ^{\xi } )=f$\textit{ is satisfied;}

\textit{ 2) when fulfilling equations of matter fields }$G_{\xi }^{} =0$\textit{ (on the extremal of matter fields):}

\textit{ a) the law of motion of matter }$\nabla _{\sigma } \tau {\kern 1pt} _{n}^{\sigma } =\tau _{m}^{\mu } F_{\mu \, n}^{m} $\textit{ is fulfilled;}

\textit{ b) under the additional assumption about conservation of the total energy-momentum }$\nabla _{\mu } T{\kern 1pt} _{m}^{\mu } =0$\textit{, equations }$(\nabla _{\nu } B_{n}^{\mu \nu } +T_{n}^{\mu } )F_{\mu m}^{n} =0$\textit{ are fulfilled, which are equivalent to the part of equations of the gravitational field.}

As we can see, this result does not depend on a specific expression for Lagrangians and is a consequence only of the gauge translational invariance of the theory of gravity.

\section{Laws of motion of charged matter}

Let us now separate from the matter fields the gauge field $A_{\mu }^{\, i} $ (indices $i, j, k$) of internal symmetry $V^{g} $ (hereinafter simply the gauge field), and by matter we mean all fields $\psi ^{\xi } $, except the gravitational and the gauge fields. We will consider the matter to be gauge charged. So the Lagrangian of the theory in this case has three components:

\[L=\sqrt{g} L_{G} (h,\partial h,...,\partial ^{(n)} h)+\sqrt{g} L_{A} (h,A,\partial A)+\sqrt{g} L_{\psi } (h,\partial h,A,\psi ,\partial \psi )\]
and the field variables $q^{I} $ split into a system $\{ \, h_{\mu }^{m} ,A{\kern 1pt} _{\nu }^{i} ,\psi ^{\xi } \} $, which is equivalent to splitting the field index $I$ into a multiindex $\{ \, _{\mu }^{m} ,\, _{\nu }^{\, i} ,{\kern 1pt} ^{\xi } \} $.

The symmetry of theories under consideration is described by the generalized gauge deformed group $G_{M}^{g} =V_{}^{g} \times )\, T_{M}^{g} $, which has the structure of the semidirect product of groups $V^{g} $ and $T_{M}^{g} $ (Appendix A.2). We will not concretize the theory by specifying the components of its Lagrangian and will consider it in general based solely on the symmetry of the theory. We consider all components of total Lagrangian $L$ to be $G_{M}^{g} $-invariant. Note only that among theories under consideration, there is a canonical variant, which is the general relativity (in the tetrad formalism) together with the Maxwell's electrodynamics  \cite{Sam ClQGr}. However, even in the case of Einstein's theory of gravity the gauge field may not satisfy the Young-Mills equations, for example, in the case of non-quadratic dependence of the Lagrangian $\sqrt{g} L_{A} $ on the tensor of the gauge field $F_{\mu \nu }^{\, i} $.

Infinitesimal transformations of field variables under the action of the group $G_{M}^{g} $ are described by formulas (\ref{eq102}), (\ref{eq115}):

\begin{equation} \label{eq52} \eqalign {{\delta {\kern 1pt} h_{\mu }^{m} =-F_{\mu {\kern 1pt} n}^{m} {\kern 1pt} t^{n} -\partial _{\mu } t^{m} ,    \qquad     \delta A_{\mu }^{{\kern 1pt} i} =F_{\mu {\kern 1pt} n}^{i} {\kern 1pt} {\kern 1pt} t^{n} -F_{jk}^{i} A_{\mu }^{j} \upsilon ^{k} -\partial _{\mu } \upsilon ^{i} , }  \cr   {\delta \psi ^{\xi } =-(\partial _{n} +A_{n}^{i} \, Z_{i} ){\kern 1pt} {\kern 1pt} \psi ^{\xi } {\kern 1pt} t^{n} +Z_{i} {\kern 1pt} {\kern 1pt} \psi ^{\xi } {\kern 1pt} {\kern 1pt} \upsilon ^{i} \,,} }                                                                                \end{equation}
from which we find following expressions for the coefficients corresponding to formula (\ref{eq16}):

\begin{equation} \label{eq53} \eqalign { {a_{\mu {\kern 1pt} n}^{m} =-F_{\mu {\kern 1pt} n}^{m} ,  \qquad   a_{\mu {\kern 1pt} n}^{\, i} =F_{\mu {\kern 1pt} n}^{\, i} ,  \qquad   a_{\mu {\kern 1pt} k}^{\, i} =-F_{j{\kern 1pt} k}^{i} A_{\mu }^{j} , } \cr {a_{\; {\kern 1pt} n}^{\xi } =-(\partial _{n} +A_{\, n}^{\, i} \, Z_{i} ){\kern 1pt} {\kern 1pt} \psi ^{\xi } ,   \qquad  a_{\; {\kern 1pt} i}^{\xi } =Z_{i} {\kern 1pt} {\kern 1pt} \psi ^{\xi },} \cr {b_{\mu {\kern 1pt} n}^{m\nu } =-\delta _{\mu }^{\nu } \delta _{n}^{m} ,  \qquad   b_{\mu {\kern 1pt} j}^{\, i \,\nu } =-\delta _{\mu }^{\nu } \delta _{j}^{i} .}}
\end{equation}
Here $\upsilon ^{i} $ are the parameters of the gauge group of internal symmetry $V^{g} $, which will be marked by the indices $i,j,k$, and $Z_{i} $ are the generators of the field representation of the group $V^{g} $.

In addition to the above notations for variational derivatives (\ref{eq28}), (\ref{eq30}), (\ref{eq31}), we introduce notations for variational derivatives of Lagrangian $\sqrt{g} L_{A} $:

\begin{equation} \label{eq54} [\sqrt{g} L_{A} ]_{\, i}^{\mu } =:\sqrt{g} G_{\, i}^{\mu } \,,\qquad      [\sqrt{g} L_{A} ]_{m}^{\mu } =:-\sqrt{g} {\kern 1pt} \theta _{m}^{\mu } ,                                                                    \end{equation}
as well as the variational derivative of the matter Lagrangian $\sqrt{g} L_{\psi } $

\begin{equation} \label{eq55} [\sqrt{g} L_{\psi } ]_{\, i}^{\mu } =:-\sqrt{g} j_{\, i}^{\mu } ,                                                                                                     \end{equation}
which appears due to the fact that matter in our case is considered to be gauge charged. With these notations, the variational derivatives of total Lagrangian $L$ with respect to the potentials $h_{\mu }^{m} $ and $A_{\mu }^{{\kern 1pt} i} $ of gauge fields of both external (space-time) and internal symmetry take the form:

\[[L]_{m}^{\mu } =\sqrt{g} (G_{m}^{\mu } -\theta _{m}^{\mu } -\tau _{m}^{\mu } ),   \qquad    [L]_{{\kern 1pt} i}^{\mu } =\sqrt{g} (G_{{\kern 1pt} i}^{\mu } -j_{{\kern 1pt} i}^{\mu } ).                       \]

The expression for the variational derivatives with respect to matter fields remains unchanged

\begin{equation} \label{eq56} [L]{\kern 1pt} _{\xi } =[\sqrt{g} L_{\psi } ]{\kern 1pt} _{\xi } =\sqrt{g} {\kern 1pt} G_{\xi } ,                                                                                                    \end{equation}
although the functions ${\kern 1pt} G_{\xi }^{} =f_{\xi } -\nabla _{\sigma } p_{\xi }^{\sigma } $ itself in the presence of the gauge field change due to the dependence of the matter Lagrangian $\sqrt{g} L_{\psi } $ on the fields $A_{\mu }^{{\kern 1pt} i} $.

From the first of definitions (\ref{eq54}) follows

\begin{equation} \label{eq57} G_{{\kern 1pt} i}^{\mu } =-i_{{\kern 1pt} i}^{\mu } -\nabla _{\sigma } B_{{\kern 1pt} i}^{\mu {\kern 1pt} \sigma } ,
\end{equation}
where

\[\, B_{\, {\kern 1pt} i}^{\mu {\kern 1pt} \sigma } :=\partial _{{\kern 1pt} i}^{\nu ,\sigma } L_{A} ,    \qquad       i_{{\kern 1pt} i}^{\mu } :=-\partial _{{\kern 1pt} i}^{\mu } L_{A} .                                        \]
When writing expression (\ref{eq57}), it is taken into account that tensor $B_{\, {\kern 1pt} i}^{\mu {\kern 1pt} \sigma } $ is a tensor superpotential associated with transformations of internal symmetry $V^{g} $, as follows from the definition (\ref{eq25}) applied to transformations (\ref{eq52}). Therefore, tensor $B_{\, {\kern 1pt} i}^{\mu {\kern 1pt} \sigma } $ is antisymmetric with respect to upper indices. Tensor $B_{\, {\kern 1pt} i}^{\mu {\kern 1pt} \sigma } $ will be called the \textit{tensor of induction of the gauge field}  $A_{\mu }^{{\kern 1pt} i} $.

In addition, from the second definition (\ref{eq54}) follows ${\kern 1pt} \theta _{m}^{\mu } =-\partial _{m}^{\mu } (\sqrt{g} L_{A} )/\sqrt{g} $, and from definition (\ref{eq55}) follows $j_{{\kern 1pt} i}^{\mu } =-\partial _{{\kern 1pt} i}^{\mu } L_{\psi } $.

With the Lagrangian of the gauge field $\sqrt{g} L_{A} $, new components of currents appear, which will be denoted by the index $A$. In addition, the expressions for the currents of the matter fields change because they become gauge charged.

Noether's currents associated with gauge translations specify formula (\ref{eq24}) taking into account expressions (\ref{eq53}) for the corresponding coefficients:

\begin{equation} \label{eq58} J_{A{\kern 1pt} {\kern 1pt} n}^{\, \; \; \nu } =\, \, -\sqrt{g} (B_{\, i}^{\mu {\kern 1pt} \nu } \, F_{\mu {\kern 1pt} n}^{{\kern 1pt} i} +L_{A} {\kern 1pt} h_{n}^{\nu } )\,,                                                                                         \end{equation}

\begin{equation} \label{eq59} J{\kern 1pt} _{\psi {\kern 1pt} n}^{\, \; \, \, \nu } =\, \, \sqrt{g} {\kern 1pt} [\beta _{m}^{\mu \, \nu } F{\kern 1pt} _{\mu {\kern 1pt} n}^{m{\kern 1pt} {\kern 1pt} } +\, p_{\xi }^{\nu } (\partial _{n} +A_{n}^{{\kern 1pt} i} Z_{i} ){\kern 1pt} {\kern 1pt} \psi ^{\xi } -L_{\psi } {\kern 1pt} h_{n}^{\nu } ]\,.
\end{equation}
For the gravitational field, the tensor density of energy-momentum remains the same (\ref{eq33}). Currents associated with the gauge transformations of the internal symmetry are given by expressions:

\[J_{A{\kern 1pt} {\kern 1pt} i}^{\, \; \; \nu } =\, \, \sqrt{g} B_{j}^{\mu {\kern 1pt} {\kern 1pt} \nu } \, F_{k{\kern 1pt} i}^{j} A_{\mu }^{k}  ,   \qquad     J{\kern 1pt} _{\psi {\kern 1pt} \, i}^{\, \; \, \, \nu } =\, -\sqrt{g} p_{\xi }^{\nu } Z_{{\kern 1pt} i} {\kern 1pt} {\kern 1pt} \psi ^{\xi } ,                                                            \]
hence

\begin{equation} \label{eq60} J{\kern 1pt} _{\psi {\kern 1pt} n}^{\, \; \, \, \nu } =\, \sqrt{g} (\beta _{m}^{\mu \, \nu } F{\kern 1pt} _{\mu {\kern 1pt} n}^{m{\kern 1pt} {\kern 1pt} } +\, p_{\xi }^{\nu } \partial _{n} \psi ^{\xi } {\kern 1pt} {\kern 1pt} -L_{\psi } {\kern 1pt} h_{n}^{\nu } )-A{\kern 1pt} _{n}^{{\kern 1pt} i} \, J{\kern 1pt} _{\psi {\kern 1pt} \, i}^{\, \; \, \, \nu } \,.                                                          \end{equation}

Identity (\ref{eq21}), applied to $T_{M}^{g} $-transformations of Lagrangian $\sqrt{g} L_{A} $, gives $\sqrt{g} {\kern 1pt} \theta _{m}^{\mu } =J_{A{\kern 1pt} {\kern 1pt} m}^{\, \; \; \mu } $ and indicates that the quantity ${\kern 1pt} \theta _{m}^{\mu } $ defined by the second relation in (\ref{eq54}) is the energy-momentum tensor of the gauge field. The application of identity (\ref{eq21}) for $T_{M}^{g} $-transformations of total Lagrangian $L$ is reduced to

\begin{equation} \label{eq61} -[L]_{m}^{\mu } /\sqrt{g} =-(G_{m}^{\mu } -\theta _{m}^{\mu } -\tau _{m}^{\mu } )=T{\kern 1pt} _{m}^{\mu } +\nabla _{\sigma } B{\kern 1pt} _{m}^{\mu \sigma } \,,                                                               \end{equation}
where $T_{\, m}^{\mu } \, =t_{m}^{\mu } +\theta _{m}^{\mu } +\tau _{m}^{\mu } $ is now the total energy-momentum tensor of the gravitational, gauge and matter fields. From this identity it follows that equations of the gravitational field $[L]_{m}^{\mu } =0$ also in this case can be written both in the form of the Einstein equation $G_{m}^{\mu } =\theta _{m}^{\mu } +\tau _{m}^{\mu } $ and in the superpotential form $\nabla _{\sigma } B{\kern 1pt} _{m}^{\mu \sigma } =-T{\kern 1pt} _{m}^{\mu } $.

Identity (\ref{eq21}) for $V_{}^{g} $-transformations of Lagrangians $\sqrt{g} L_{A} $ and $\sqrt{g} L_{\psi } $ gives:

\begin{equation} \label{eq62} -G_{i}^{\nu } =J_{A\,i}^{\, \; \, \, \nu } /\sqrt{g} +\, \nabla _{\sigma } B_{\, i}^{\nu \sigma } \,,   \qquad j_{i}^{\nu } =J _{\psi \, i}^{\, \; \, \, \nu } /\sqrt{g} \,.
\end{equation}
Comparing (\ref{eq62}) with (\ref{eq57}) we obtain $i_{{\kern 1pt} i}^{\nu } =J{\kern 1pt} _{A{\kern 1pt} {\kern 1pt} {\kern 1pt} i}^{\, \; \, \, \nu } /\sqrt{g} $. Therefore, the vectors $i_{{\kern 1pt} i}^{\nu } $ and $j_{{\kern 1pt} i}^{\nu } $ are the currents of the gauge charges of gauge and matter fields, respectively.

Due to the $V_{}^{g} $-symmetry of total Lagrangian $L$, identity (\ref{eq21}) is reduced to

\begin{equation} \label{eq63} -[L]_{{\kern 1pt} i}^{\nu } /\sqrt{g} =-\, (G_{{\kern 1pt} i}^{\nu } -j_{{\kern 1pt} i}^{\nu } )=I_{{\kern 1pt} {\kern 1pt} i}^{\, \nu } +\nabla _{\sigma } B_{\, {\kern 1pt} i}^{\nu {\kern 1pt} \sigma } \,,                                                               \end{equation}
where $I_{{\kern 1pt} i}^{\nu } =i_{{\kern 1pt} i}^{\nu } +j_{{\kern 1pt} i}^{\nu } $ is the total current of the gauge charges of the gauge field and the fields of matter. Therefore, according to Theorem (Section 2) equations of the gauge field $[L]_{{\kern 1pt} i}^{\nu } =0$, as well as equations of the gravitational field, can be written both in the form of the Einstein equation $G_{{\kern 1pt} i}^{\nu } =j_{{\kern 1pt} i}^{\nu } $ and in the superpotential form

\begin{equation} \label{eq64} \nabla _{\sigma } B_{\, {\kern 1pt} i}^{\nu {\kern 1pt} \sigma } =-I_{{\kern 1pt} {\kern 1pt} i}^{\nu }\, ,                                                                                                                    \end{equation}
that is, in the form of Yang-Mills equations, which is an expression of the total current of the gauge charges $I_{{\kern 1pt} {\kern 1pt} i}^{\nu } $ through the covariant divergence of the tensor of induction of the gauge field $B_{\, {\kern 1pt} i}^{\nu {\kern 1pt} \sigma } $, which is its tensor superpotential. This is a consequence of the $V_{}^{g} $-invariance of the theory and does not depend on the specific type of Lagrangians $\sqrt{g} L_{A} $ and $\sqrt{g} L_{\psi } $, which define specific expressions for the quantities $B_{\, {\kern 1pt} i}^{\nu {\kern 1pt} \sigma } $ and $I_{{\kern 1pt} {\kern 1pt} i}^{\nu } $. Tensor $B_{\, {\kern 1pt} i}^{\nu {\kern 1pt} \sigma } $ is \textit{the power characteristic of the gauge field}.

Identities (\ref{eq20}) corresponding to the $T_{M}^{g} $-invariance of Lagrangians $\sqrt{g} L_{G} $, $\sqrt{g} L_{A} $ and $\sqrt{g} L_{\psi } $ have the following form:

\begin{equation} \label{eq65} -G_{m}^{\mu } {\kern 1pt} {\kern 1pt} F{\kern 1pt} _{\mu \, n}^{m{\kern 1pt} {\kern 1pt} } =\nabla _{\sigma } t{\kern 1pt} _{n}^{\sigma }\,,                                                                                                             \end{equation}

\begin{equation} \label{eq66} \theta _{m}^{\mu } {\kern 1pt} {\kern 1pt} F{\kern 1pt} _{\mu \, n}^{m{\kern 1pt} {\kern 1pt} } +G_{\, i}^{\mu } {\kern 1pt} F_{\mu {\kern 1pt} n}^{\, i} =\nabla _{\sigma } \theta _{n}^{\sigma } \,,                                                                                                \end{equation}

\begin{equation} \label{eq67} \tau _{m}^{\mu } {\kern 1pt} F{\kern 1pt} _{\mu \, n}^{m{\kern 1pt} {\kern 1pt} } -j_{\, i}^{\mu } {\kern 1pt} {\kern 1pt} F_{\mu {\kern 1pt} n}^{\, i} -G_{\xi } (\partial _{n} +A_{n}^{\, i} Z_{\, i} ){\kern 1pt} {\kern 1pt} \psi ^{\xi } {\kern 1pt} =\nabla _{\sigma } \tau {\kern 1pt} _{n}^{\sigma }  \end{equation}
and in the sum give an identity corresponding to total Lagrangian $L$:

\begin{equation} \label{eq68} -(G_{m}^{\mu } -\theta _{m}^{\mu } -\tau _{m}^{\mu } ){\kern 1pt} F{\kern 1pt} _{\mu \, n}^{m{\kern 1pt} {\kern 1pt} } +(G_{\, i}^{\mu } -j_{\, i}^{\mu } {\kern 1pt} ){\kern 1pt} F_{\mu {\kern 1pt} n}^{\, i} -G_{\xi } (\partial _{n} +A_{n}^{\, i} Z_{\, i} ){\kern 1pt} {\kern 1pt} \psi ^{\xi } {\kern 1pt} =\nabla _{\sigma } T{\kern 1pt} _{n}^{\sigma } .                             \end{equation}

We now write identities (\ref{eq20}) corresponding to the $V_{}^{g} $-invariance of Lagrangians $\sqrt{g} L_{A} $ and $\sqrt{g} L_{\psi } $:

\begin{equation} \label{eq69} -G_{j}^{\mu } F_{{\kern 1pt} k{\kern 1pt} i}^{j} A_{\mu }^{k} =\nabla _{\sigma } i{\kern 1pt} _{\, i}^{\sigma }\,,  \end{equation}

\begin{equation} \label{eq70} j_{j}^{\mu } F_{{\kern 1pt} k{\kern 1pt} i}^{j} A_{\mu }^{k} +G_{\xi } {\kern 1pt} Z_{\, i} {\kern 1pt} {\kern 1pt} \psi ^{\xi } =\nabla _{\sigma } j{\kern 1pt} _{i}^{\sigma } \,,                                                                                           \end{equation}
as well as their sum:

\begin{equation} \label{eq71} -(G_{j}^{\mu } -j_{j}^{\mu } )F_{k{\kern 1pt} i}^{j} A_{\mu }^{k} +G_{\xi } {\kern 1pt} Z_{{\kern 1pt} i} {\kern 1pt} {\kern 1pt} \psi ^{\xi } =\nabla _{\sigma } I{\kern 1pt} _{i}^{\sigma } \,,                                                                             \end{equation}
where it follows

\begin{equation} \label{eq72} G_{\xi } {\kern 1pt} Z_{{\kern 1pt} i} {\kern 1pt} {\kern 1pt} \psi ^{\xi } =\nabla _{\sigma } I{\kern 1pt} _{i}^{\sigma } +(G_{j}^{\mu } -j_{j}^{\mu } )F_{{\kern 1pt} k{\kern 1pt} i}^{j} A_{\mu }^{k} \,,                                                                                \end{equation}
which allows identity (\ref{eq67}) to give another form

\begin{equation} \label{eq73} \nabla _{\sigma } \tau {\kern 1pt} _{n}^{\sigma } {\kern 1pt} =\tau _{m}^{\mu } {\kern 1pt} F{\kern 1pt} _{\mu \, n}^{m{\kern 1pt} {\kern 1pt} } -j_{{\kern 1pt} i}^{\mu } {\kern 1pt} {\kern 1pt} F_{\mu {\kern 1pt} n}^{{\kern 1pt} i} -\nabla _{\sigma } I{\kern 1pt} _{i}^{\sigma } A_{n}^{{\kern 1pt} i} -(G_{j}^{\mu } -j_{j}^{\mu } )F_{{\kern 1pt} k{\kern 1pt} i}^{j} A_{\mu }^{k} A_{n}^{{\kern 1pt} i} -G_{\xi } {\kern 1pt} \partial _{n} {\kern 1pt} \psi ^{\xi } \,,                           \end{equation}
and identities (\ref{eq70}) to give the form:

\begin{equation} \label{eq74} \nabla _{\sigma } j{\kern 1pt} _{i}^{\sigma } =j_{j}^{\mu } F_{{\kern 1pt} k{\kern 1pt} i}^{j} A_{\mu }^{k} +\nabla _{\sigma } I{\kern 1pt} _{i}^{\sigma } +(G_{j}^{\mu } -j_{j}^{\mu } )F_{{\kern 1pt} k{\kern 1pt} i}^{j} A_{\mu }^{k} \,.                                                                  \end{equation}

Until now, in this section, we have considered exceptionally strong identities. Let's consider what these identities on extremals lead to (weakened version of identities).

 On the gravitational extremal

\begin{equation} \label{eq75} -[L]_{m}^{\mu } /\sqrt{g} =-(G_{m}^{\mu } -\theta _{m}^{\mu } -\tau _{m}^{\mu } )=T{\kern 1pt} _{m}^{\mu } +\nabla _{\sigma } B{\kern 1pt} _{m}^{\mu \sigma } =0,   \quad   \nabla _{\sigma } T{\kern 1pt} _{n}^{\sigma } =0\,,
\end{equation}
so from identity (\ref{eq68}) follows

\begin{equation} \label{eq76} (G_{\, i}^{\mu } -j_{\, i}^{\mu } {\kern 1pt} ){\kern 1pt} F_{\mu {\kern 1pt} n}^{\, i} =G_{\xi } (\partial _{n} +A_{n}^{\, i} Z_{\, i} ){\kern 1pt} {\kern 1pt} \psi ^{\xi } \,,                                                                                  \end{equation}
which, taking into account identity (\ref{eq67}) gives:

\begin{equation} \label{eq77} \nabla _{\sigma } \tau _{n}^{\sigma } =\tau _{m}^{\mu } {\kern 1pt} {\kern 1pt} F{\kern 1pt} _{\mu \, n}^{m{\kern 1pt} {\kern 1pt} } -j_{{\kern 1pt} i}^{\mu } F_{\mu {\kern 1pt} n}^{{\kern 1pt} i} -(G_{{\kern 1pt} i}^{\mu } {\kern 1pt} -j_{{\kern 1pt} i}^{\mu } )F_{\mu {\kern 1pt} n}^{{\kern 1pt} i} \,.                                                                        \end{equation}

On the extremal of gauge field

\begin{equation} \label{eq78} -[L]_{{\kern 1pt} i}^{\nu } /\sqrt{g} =-\, (G_{{\kern 1pt} i}^{\nu } -j_{{\kern 1pt} i}^{\nu } )=I_{{\kern 1pt} {\kern 1pt} i}^{\, \nu } +\nabla _{\sigma } B_{\, {\kern 1pt} i}^{\nu {\kern 1pt} \sigma } =0,      \quad \nabla _{\sigma } I{\kern 1pt} _{{\kern 1pt} i}^{\sigma } =0 \,,
\end{equation}
so from (\ref{eq71}) follows:

\begin{equation} \label{eq79} G_{\xi } {\kern 1pt} Z_{{\kern 1pt} i} {\kern 1pt} {\kern 1pt} \psi ^{\xi } =0 \end{equation}
and from (\ref{eq74}) follows

\begin{equation} \label{eq80} \nabla _{\sigma } j{\kern 1pt} _{{\kern 1pt} i}^{\sigma } =j_{j}^{\mu } F_{{\kern 1pt} k{\kern 1pt} {\kern 1pt} i}^{j} A_{\mu }^{k} \,, \end{equation}
which is the \textit{equation of transfer of gauge charges of matter}. Identity (\ref{eq66}) is reduced to identity
\begin{equation} \label{eq81} \nabla _{\sigma } \theta _{n}^{\sigma } =\theta _{m}^{\mu } {\kern 1pt} {\kern 1pt} F_{\mu \, n}^{m{\kern 1pt} {\kern 1pt} } +j_{{\kern 1pt} i}^{\mu } {\kern 1pt} F_{\mu {\kern 1pt} n}^{{\kern 1pt} i} \,, \end{equation}
which is the \textit{equation of transfer of energy-momentum of the gauge field}. Identity (\ref{eq67}), taking into account (\ref{eq79}), is reduced to identity

\begin{equation} \label{eq82} \nabla _{\sigma } \tau {\kern 1pt} _{n}^{\sigma } {\kern 1pt} =\tau _{m}^{\mu } {\kern 1pt} F{\kern 1pt} _{\mu \, n}^{m{\kern 1pt} {\kern 1pt} } -j_{{\kern 1pt} i}^{\mu } {\kern 1pt} {\kern 1pt} F_{\mu {\kern 1pt} n}^{{\kern 1pt} i} -G_{\xi } {\kern 1pt} \partial _{n} {\kern 1pt} \psi ^{\xi } \,,                                                                                \end{equation}
and identity (\ref{eq68}) is reduced to identity

\begin{equation} \label{eq83} -(G_{m}^{\mu } -\theta _{m}^{\mu } -\tau _{m}^{\mu } ){\kern 1pt} F_{\mu \, n}^{m{\kern 1pt} {\kern 1pt} } -G_{\xi } {\kern 1pt} \partial _{n} {\kern 1pt} {\kern 1pt} \psi ^{\xi } {\kern 1pt} =\nabla _{\sigma } T{\kern 1pt} _{n}^{\sigma } \,.                                                                        \end{equation}

On the extremal of matter fields $[L]{\kern 1pt} _{\xi }^{} =\sqrt{g} G_{\xi }^{} =0$, so from (\ref{eq67}) follows\textbf{\textit{}}

\begin{equation} \label{eq84} \nabla _{\sigma } \tau {\kern 1pt} _{n}^{\sigma } =\tau _{m}^{\mu } {\kern 1pt} F{\kern 1pt} _{\mu \, n}^{m{\kern 1pt} {\kern 1pt} } -j_{{\kern 1pt} i}^{\mu } {\kern 1pt} {\kern 1pt} F_{\mu {\kern 1pt} n}^{{\kern 1pt} i} \,, \end{equation}
which is \textit{the equation of transfer of energy-momentum of gauge charged matter, }or\textit{ the law of its motion}. The presence of gauge interaction is described here by an additional (in comparison with formula (\ref{eq46})) term $-j_{\, i}^{\mu } {\kern 1pt} {\kern 1pt} F_{\mu {\kern 1pt} n}^{\, i} {\kern 1pt} $, which is the Lorentz force density generalized to the case of an arbitrary gauge field of internal symmetry. In addition, from identity (\ref{eq68}) follows identity

\begin{equation} \label{eq85} -(G_{m}^{\mu } -\theta _{m}^{\mu } -\tau _{m}^{\mu } ){\kern 1pt} F{\kern 1pt} _{\mu \, n}^{m{\kern 1pt} {\kern 1pt} } +(G_{{\kern 1pt} i}^{\mu } -j_{{\kern 1pt} i}^{\mu } {\kern 1pt} ){\kern 1pt} F_{\mu {\kern 1pt} n}^{{\kern 1pt} i} {\kern 1pt} =\nabla _{\sigma } T{\kern 1pt} _{n}^{\sigma } \,,
\end{equation}
and from (\ref{eq70}) again follows the equation of transfer of gauge charge of matter (\ref{eq80}).

From (\ref{eq84}) follows that as tensor $F{\kern 1pt} _{\mu \, n}^{m{\kern 1pt} {\kern 1pt} } $ describes strength of gravitational field and $\tau _{m}^{\mu } {\kern 1pt} $ the gravitational charge of matter, tensor $F_{\mu {\kern 1pt} n}^{\, i} $ describes \textit{strength of the gauge field} and $j_{\, i}^{\mu } $  the \textit{gauge charge of matter}.\textbf{}

At the simultaneous fulfilling of the equations of gravitational and gauge fields, from identity (\ref{eq77}) we obtain the law of motion of gauge charged matter (\ref{eq84}). At the same time, from identity (\ref{eq83}) follow equations $G_{\xi } \, \partial _{n} {\kern 1pt} {\kern 1pt} \psi ^{\xi } {\kern 1pt} =0$, which generalize equations (\ref{eq49}) in the case of gauge interactions (which is taken into account in our case in expression (\ref{eq56}) for $G_{\xi } $).

When equations of the gravitational field and fields of matter are fulfilled simultaneously, identity (\ref{eq85}) (or (\ref{eq77})) is reduced to equations

\[(G_{{\kern 1pt} i}^{\mu } -j_{{\kern 1pt} i}^{\mu } {\kern 1pt} ){\kern 1pt} F_{\mu {\kern 1pt} n}^{{\kern 1pt} i} {\kern 1pt} =-(I_{{\kern 1pt} {\kern 1pt} i}^{\, \mu } +\nabla _{\sigma } B_{\, {\kern 1pt} i}^{\mu {\kern 1pt} \sigma } )F_{\mu {\kern 1pt} n}^{{\kern 1pt} i} =0,                                \]
which are equivalent to the part of the gauge field equations.

When equations of the gauge field and the fields of matter are fulfilled simultaneously, identity (\ref{eq83}) gives

\[-(G_{m}^{\mu } -\theta _{m}^{\mu } -\tau _{m}^{\mu } ){\kern 1pt} F_{\mu \, n}^{m{\kern 1pt} {\kern 1pt} } {\kern 1pt} =\nabla _{\sigma } T{\kern 1pt} _{n}^{\sigma } ,                                          \]
and under the condition of an additional assumption about the conservation of the total energy-momentum $\nabla _{\sigma } T{\kern 1pt} _{n}^{\sigma } =0$, taking into account (\ref{eq61}), we obtain the identity

\[(\nabla _{\nu } B_{n}^{\mu \nu } +T_{n}^{\mu } )F_{\mu m}^{n} =0,                                       \]
which generalizes equations (\ref{eq51}) in the case of the presence of a gauge field, which is taken into account by the term $\theta _{m}^{\mu } $ in the expression for $T_{m}^{\mu } $.

We summarize the obtained results.

\textbf{ Proposition 2.} \textit{In any gauge theory of the generalized gauge group} $G_{M}^{g} =V_{}^{g} \times )\, T_{M}^{g}$:

\textit{ 1) when fulfilling equations of the gravitational field $\nabla _{\sigma } B{\kern 1pt} _{m}^{\mu \sigma } =-T{\kern 1pt} _{m}^{\mu } $ and the gauge field of internal symmetry }$\nabla _{\sigma } B_{\, {\kern 1pt} i}^{\mu {\kern 1pt} \sigma } =-I_{{\kern 1pt} {\kern 1pt} i}^{\mu } $ \textit{ (on the gravitational and gauge extremals):}

\textit{ a) the law of motion of gauge charged matter }$\nabla _{\sigma } \tau {\kern 1pt} _{n}^{\sigma } =\tau _{m}^{\mu } {\kern 1pt} F{\kern 1pt} _{\mu \, n}^{m{\kern 1pt} {\kern 1pt} } -j_{\, i}^{\mu } {\kern 1pt} {\kern 1pt} F_{\mu {\kern 1pt} n}^{\, i} {\kern 1pt} $ \textit{is fulfilled;}

\textit{ b) equations }$G_{\xi }^{} \partial _{n} {\kern 1pt} \psi ^{\xi } =0$\textit{ are fulfilled, which are equivalent to equations of matter fields }$G_{\xi }^{} =0$\textit{ if the condition }$rank{\kern 1pt} {\kern 1pt} (\partial _{n} {\kern 1pt} \psi ^{\xi } )=f$\textit{ is satisfied;}

\textit{ 2) when fulfilling equations of the gauge field }$\nabla _{\sigma } B_{\, {\kern 1pt} i}^{\mu {\kern 1pt} \sigma } =-I_{{\kern 1pt} {\kern 1pt} i}^{\mu } $\textit{ (on the extremal of gauge field):}

\textit{ a) the equation of transfer of gauge charges of matter }$\nabla _{\sigma } j{\kern 1pt} _{{\kern 1pt} i}^{\sigma } =j_{j}^{\mu } F_{{\kern 1pt} k{\kern 1pt} {\kern 1pt} i}^{j} A_{\mu }^{k} $ \textit{is fulfilled;}

\textit{ b) the equation of transfer of energy-momentum of the gauge field }$\nabla _{\sigma } \theta _{n}^{\sigma } =\theta _{m}^{\mu } {\kern 1pt} {\kern 1pt} F_{\mu \, n}^{m{\kern 1pt} {\kern 1pt} } +j_{{\kern 1pt} i}^{\mu } {\kern 1pt} F_{\mu {\kern 1pt} n}^{{\kern 1pt} i} $ \textit{is fulfilled;}

\textit{ 3) when fulfilling equations of matter fields }$G_{\xi }^{} =0$\textit{ (on the extremal of the matter fields):}

\textit{ a) the law of motion of gauge charged matter }$\nabla _{\sigma } \tau {\kern 1pt} _{n}^{\sigma } =\tau _{m}^{\mu } {\kern 1pt} F{\kern 1pt} _{\mu \, n}^{m{\kern 1pt} {\kern 1pt} } -j_{\, i}^{\mu } {\kern 1pt} {\kern 1pt} F_{\mu {\kern 1pt} n}^{\, i} {\kern 1pt} $ \textit{is fulfilled;}

\textit{ b) under the additional assumption of the fulfilling of equations of the gravitational field $\nabla _{\sigma } B{\kern 1pt} _{m}^{\mu \sigma } =-T{\kern 1pt} _{m}^{\mu } $, equations }$(I_{{\kern 1pt} {\kern 1pt} i}^{\, \mu } +\nabla _{\sigma } B_{\, {\kern 1pt} i}^{\mu {\kern 1pt} \sigma } )F_{\mu {\kern 1pt} n}^{{\kern 1pt} i} =0$ \textit{ are fulfilled, which are equivalent to the part of equations of the gauge field;}

\textit{ c) under the additional assumption of the fulfilling of equations of the gauge field }$\nabla _{\sigma } B_{\, {\kern 1pt} i}^{\mu {\kern 1pt} \sigma } =-I_{{\kern 1pt} {\kern 1pt} i}^{\mu } $\textit{, as well as the assumption about conservation of the total energy-momentum }$\nabla _{\mu } T{\kern 1pt} _{m}^{\mu } =0$\textit{, equations }$(\nabla _{\nu } B_{n}^{\mu \nu } +T_{n}^{\mu } )F_{\mu m}^{n} =0$\textit{ are fulfilled, which are equivalent to the part of equations of the gravitational field.}

We emphasize once again that these results do not depend on specific expressions for Lagrangians and are a consequence only of the gauge invariance of the theory.

Point 3b) of this Proposition generalizes the result of Hilbert \cite{Hilbert} to the arbitrary gauge field and the presence of currents of matter charges.

Generalized gauge groups $G_M^g=V^g \left. \times \right)\, T_M^g$ have no additional parameters, other than parameters of groups $V^g$ and $T_M^g$, and therefore have no additional conservation laws (other than those discussed in the Proposition 2).

Let's deform the group $G_{M}^{g} =U(1)^{g} \left. \times \right)\, T_{M}^{g} $ by changing the method of formation of the semidirect product so that at the infinitesimal level it reduced to reparameterization of the group $G_{M}^{g} $, namely, instead of $\upsilon ^{{\kern 1pt} i} $ we introduce new parameters $\bar{\upsilon }^{{\kern 1pt} i} $ according to the formula $\upsilon ^{{\kern 1pt} i} =\bar{\upsilon }^{{\kern 1pt} i} +C_{n}^{{\kern 1pt} i} \, t^{n} $ with arbitrary deformation coefficients $C_{n}^{{\kern 1pt} i} \, $ (Appendix). In this case, part of transformations (\ref{eq52}) of the group $G_{M}^{g} $ changes:

\[\delta A_{\mu }^{{\kern 1pt} i} =(F_{\mu {\kern 1pt} n}^{{\kern 1pt} i} {\kern 1pt} {\kern 1pt} -F_{j{\kern 1pt} k}^{{\kern 1pt} i} A_{\mu }^{j} C_{n}^{k} -\partial _{\mu } C_{n}^{{\kern 1pt} i} )\, t^{n} -C_{n}^{{\kern 1pt} i} \partial _{\mu } {\kern 1pt} t^{n} -F_{j{\kern 1pt} k}^{{\kern 1pt} i} A_{\mu }^{j} \bar{\upsilon }^{k} -\partial _{\mu } \bar{\upsilon }^{{\kern 1pt} i}, \]

\[\delta \psi ^{\xi } =-\, [\partial _{n} +(A_{n}^{{\kern 1pt} i} -C_{n{\kern 1pt} }^{{\kern 1pt} i} )\, Z_{{\kern 1pt} i} ]{\kern 1pt} {\kern 1pt} \psi ^{\xi } {\kern 1pt} t^{n} +Z_{{\kern 1pt} i} {\kern 1pt} {\kern 1pt} \psi ^{\xi } {\kern 1pt} {\kern 1pt} \bar{\upsilon }^{{\kern 1pt} i} ,                                     \]
therefore, some of coefficients (\ref{eq53}) change, namely:

\[a_{\mu {\kern 1pt} n}^{{\kern 1pt} i} =F_{\mu {\kern 1pt} n}^{{\kern 1pt} i} -F_{j{\kern 1pt} k}^{{\kern 1pt} i} A_{\mu }^{j} C_{n}^{k} -\partial _{\mu } C_{n}^{{\kern 1pt} i} , \qquad a_{\; {\kern 1pt} n}^{\xi } =-[\partial _{n} +(A_{n}^{{\kern 1pt} i} -C_{n}^{{\kern 1pt} i} )\, Z_{{\kern 1pt} i} ]{\kern 1pt} {\kern 1pt} {\kern 1pt} \psi ^{\xi } , \]

\[ b_{\mu {\kern 1pt} n}^{{\kern 1pt} i\nu } =-\delta _{\mu }^{\nu } C_{n}^{{\kern 1pt} i} .    \]

This results in a new component of the superpotential $S_{A\, n}^{\; \; \nu {\kern 1pt} \sigma } =\sqrt{g} B_{\, {\kern 1pt} i}^{\nu {\kern 1pt} \sigma } C_{n}^{{\kern 1pt} i} $ (all others remain unchanged). In addition, the expressions for the currents associated with the translations change: \textit{there is a renormalization} of tensor densities of the energy-momentum of gauge (\ref{eq58}) and matter (\ref{eq60}) fields (energy-momentum of the gravitational field, obviously remains unchanged):

\[ {J_{A{\kern 1pt} {\kern 1pt} n}^{\, \; \; \nu } =\, -\sqrt{g} B_{{\kern 1pt} i}^{\mu {\kern 1pt} \nu } \, (F_{\mu {\kern 1pt} n}^{{\kern 1pt} i} -F_{j{\kern 1pt} k}^{{\kern 1pt} i} A_{\mu }^{j} C_{n}^{k} -\partial _{\mu } C_{n}^{{\kern 1pt} i} )-\sqrt{g} L_{A} {\kern 1pt} h_{n}^{\nu } }   \]

\[\quad \quad =\sqrt{g} (\theta _{{\kern 1pt} n}^{\nu } +i_{{\kern 1pt} {\kern 1pt} i}^{\nu } C_{n}^{{\kern 1pt} i} +B_{{\kern 1pt} i}^{\mu {\kern 1pt} \nu } \partial _{\mu } C_{n}^{{\kern 1pt} i} ) =\sqrt{g} {\kern 1pt} \theta _{{\kern 1pt} n}^{\nu } -\partial _{\mu } S_{A\, n}^{\; \; \nu {\kern 1pt} \mu }\,,
\]

\[J{\kern 1pt} _{\psi {\kern 1pt} n}^{\, \; \, \, \nu } =\, \sqrt{g} \, [\beta _{m}^{\mu \, \nu } F{\kern 1pt} _{\mu {\kern 1pt} n}^{m{\kern 1pt} {\kern 1pt} } +\, p_{\xi }^{\nu } \partial _{n} \psi ^{\xi } {\kern 1pt} {\kern 1pt} -L_{\psi } {\kern 1pt} h_{n}^{\nu } -(A{\kern 1pt} _{n}^{{\kern 1pt} i} -C{\kern 1pt} _{n}^{{\kern 1pt} i} )j_{{\kern 1pt} i}^{\nu } ]\]

\[\quad\quad =\sqrt{g} (\sigma {\kern 1pt} _{n}^{\nu } +j_{{\kern 1pt} i}^{\nu } C{\kern 1pt} _{n}^{{\kern 1pt} i} ),
\]
which leads to renormalization of the total tensor density of energy-momentum of all fields:

\[J_{{\kern 1pt} n}^{\nu } =\sqrt{g} (T_{{\kern 1pt} n}^{\nu } -\nabla _{\sigma } \beta _{\, n}^{\nu {\kern 1pt} \sigma } +I_{{\kern 1pt} {\kern 1pt} i}^{\nu } C_{n}^{{\kern 1pt} i} +B_{{\kern 1pt} i}^{\mu {\kern 1pt} \nu } \partial _{\mu } C_{n}^{{\kern 1pt} i} ),              \]
It corresponds the renormalized energy-momentum tensor $\bar{T}_{{\kern 1pt} n}^{\nu } :=J_{{\kern 1pt} n}^{\nu } /\sqrt{g} $. Taking the divergence from this quantity, we obtain:

\begin{equation} \label{eq86} \nabla _{\nu } \bar{T}_{{\kern 1pt} n}^{\nu } =\nabla _{\nu } T_{{\kern 1pt} n}^{\nu } +\nabla _{\nu } I_{{\kern 1pt} {\kern 1pt} i}^{\nu } C_{n}^{{\kern 1pt} i} +(I_{{\kern 1pt} {\kern 1pt} i}^{\nu } +\nabla _{\sigma } B_{{\kern 1pt} i}^{\nu {\kern 1pt} \sigma } )\, \partial _{\nu } C_{n}^{{\kern 1pt} i} \,.
\end{equation}
It follows that when fulfilling equations of the gravitational field which ensure the equality $\nabla _{\nu } T_{{\kern 1pt} n}^{\nu } =0$, equation (\ref{eq86}) is reduced to equation:

\begin{equation} \label{eq87} (I_{{\kern 1pt} {\kern 1pt} i}^{\nu } +\nabla _{\sigma } B_{{\kern 1pt} i}^{\nu {\kern 1pt} \sigma } )\, \partial _{\nu } C_{n}^{{\kern 1pt} i} =0 \,,                                                                                                    \end{equation}
provided that the gauge charges and the total energy-momentum of all fields are conserved (i.e. $\nabla _{\nu } I_{{\kern 1pt} {\kern 1pt} i}^{\nu } =0$ and $\nabla _{\nu } \bar{T}_{{\kern 1pt} n}^{\nu } =0$) for all methods of formation of semidirect product of subgroups $V_{}^{g} $ and $T_{M}^{g} $ in the group $G_{M}^{g} $.

The gauge group of electrodynamics $V_{}^{g} =U(1)^{g} $ is one-parameter, so equation (\ref{eq87}), in the case of arbitrary functions $C_{n}^{{\kern 1pt} } $, is reduced to Maxwell's dynamic equations $\nabla _{\sigma } B_{{\kern 1pt} }^{\nu {\kern 1pt} \sigma } =-I_{{\kern 1pt} {\kern 1pt} }^{\nu } $.  Therefore, in any $G_{M}^{g} =U(1)^{g} \times )\, T_{M}^{g} $-symmetric theory of electrogravity, Maxwell's dynamic equations $\nabla _{\sigma } B_{{\kern 1pt} }^{\nu {\kern 1pt} \sigma } =-I_{{\kern 1pt} {\kern 1pt} }^{\nu } $  and hence the law of motion of electrically charged matter $\nabla _{\sigma } \tau {\kern 1pt} _{n}^{\sigma } =\tau _{m}^{\mu } {\kern 1pt} F{\kern 1pt} _{\mu \, n}^{m{\kern 1pt} {\kern 1pt} } -j_{\, }^{\mu } {\kern 1pt} {\kern 1pt} F_{\mu {\kern 1pt} n}^{\, } {\kern 1pt} $ are fulfilled when fulfilled equations of the gravitational field $\nabla _{\sigma } B{\kern 1pt} _{m}^{\mu \sigma } =-T{\kern 1pt} _{m}^{\mu } $, as well as the conservation laws of electric charges $\nabla _{\nu } I_{{\kern 1pt} {\kern 1pt} }^{\nu } =0$ and of total energy-momentum $\nabla _{\nu } \bar{T}_{{\kern 1pt} n}^{\nu } =0$ for all methods of formation of semidirect product of subgroups $U(1)^{g} $ and $T_{M}^{g} $ in group $G_{M}^{g} $ \cite{Sam}.

\section{Discussion and conclusion }

Why do the laws of such various physical phenomena as gravity, gauge interactions of internal symmetry and, under certain conditions, the laws of motion of matter follow from each other? The cause of this phenomenon, as we see, is the infinite gauge symmetry of the theory, but not a specific type of Lagrangians. This symmetry determines the structure of field equations, for example, leads to antisymmetry of tensors of induction of gravitational and gauge fields, allows to represent field equations in superpotential form (in the form of dynamic Maxwell's equations), gives expression of conserved quantity through certain derivatives of Lagrangians, which are in field equations.  All this allows obtain equations of some fields by requiring equations of other fields and postulating conservation of certain physical quantities.

From a mathematical point of view, with a fairly wide group of infinite transformations and certain properties of coefficients $a_{a}^{I} $, which project the group parameters on variations of fields, requirement  symmetry of theory is sufficient to obtain some field equations and the principle of least action becomes redundant for it.

An important result of this article is that the motion of particles along geodesics of Riemannian space is inherent in an extremely wide range of theories of gravity and is a consequence of the gauge translational invariance of these theories under the condition of fulfilling equations of gravitational field. It is also interesting to note that the origin of the Lorentz force, generalized for gauge-charged matter, is a consequence of the gauge symmetry of the theory under the condition of fulfilling the equations of gravitational and gauge fields.

In the orthonormal frames, the strength of gravitational field in all the above theories of gravity is determined by the anholonomic coefficients $F_{\mu \nu}^{m}$, because they are exactly what forces particles to move along the geodesics. The tensor of induction of the gravitational field  $B_{m}^{\mu \nu}$, which is determined by a specific type of the theory Lagrangian, is its power characteristic, because it is born by the energy-momentum of the gravitational system. The formulas given in this paper make it easy to find tensors of induction of the gravitational field  $B_{m}^{\mu \nu}$ and corresponding energy-momentum tensors for an arbitrary gauge theory of the generalized gauge translations group $T_{M}^{g}$. In this paper we show that such theories include, in particular, $f(T)$-theories and $f(\Re)$-theories ($f(R)$-theories, Lovelock gravity, Einstein-Gauss-Bonnet gravity, etc.). Here, the expression of the tensor of induction of the gravitational field $B_{m}^{\mu \nu}$ for $f(\Re)$-theories in an orthonormal frame is given, and for GR in an arbitrary anholonomic affine frame are presented in \cite{JGSP} (for dilaton gravity, unimodular gravity, etc.). For other theories, we plan to do so in the future. However, \textit{the main provisions of this work relating to the problem of motion do not depend on specific expressions for Lagrangians, and hence specific expressions for $B_{m}^{\mu \nu}$, and are determined exclusively by the generalized gauge group of symmetry of the theory}. This, in our opinion, is the main result of this work.

\ack{The author is grateful to Alisa Gryshchenko for her help with the prepare of article.}

\appendix

\section{Group-theoretic description of gauge fields}

An infinite (local gauge) symmetry lies in a basis of modern theories of fundamental interactions. Theory of gravity (general relativity) is based on the idea of covariance with respect to the group of space-time diffeomorphisms. Theories of strong and electroweak interactions are gauge theories of internal symmetry. Moreover, the existence of these interactions is considered to be necessary for ensuring the local gauge symmetries.

But any physical theory can be written in covariant form without introduction of a gravitational field. Similarly, as first had been emphasized in \cite{1}, for any theory with global internal symmetry $G$ corresponding gauge symmetry $G^{g} $ can be ensured without introduction of nontrivial gauge fields by pure gauging. Presence of the gravitational or the gauge field of internal symmetry is manifested in presence of deformation -- curvature of Riemannian space, or fiber bundles with connection accordingly.

Formally, nontrivial gauge fields are entered by continuation of derivatives up to covariant derivatives $\partial _{\mu } \to \nabla _{\mu } $. Their commutators characterize strength of a field, which is considered, and from the geometrical point of view - curvature of corresponding space. On the other hand, covariant derivatives set infinitesimal space-time translations in the gauge field (curved spaces). That is why it is possible to suppose, that for introduction of nontrivial arbitrary gauge fields it is necessary to consider groups (wider than gauge groups $G^{g} $) which would generalize gauge groups $G^{g} $ to a case of \textit{nontrivial action on space-time manifold and} \textit{contain the information about arbitrary gauge fields} in which motion occurs. Consequently, such groups must \textit{contain the information about appropriate geometrical structures with arbitrary variable curvature} and set these geometrical structures on manifolds where they act. Hence from the mathematical point of view such groups should realize the \textit{Klein's Erlangen Program} \cite{2} for these geometrical structures.

For a long time it was considered that such groups do not exist. \textit{E.Cartan} \cite{3} has named the situation in question as \textit{Riemann-Klein's} \textit{antagonism} -- antagonism between \textit{Riemann's} and \textit{Klein's} approaches to geometry. There are attempts of modifying of the Klein's Program for geometrical structures with arbitrary variable curvature by means of refusal of group structure of used transformations with usage of categories \cite{4} quasigroups \cite{5} and so on. One can encounter widespread opinion that \textit{nonassociativity} is an algebraic equivalent of the geometrical notion of a curvature \cite{5}.

In this Appendix we shall show that \textit{realization of the Klein's Program for the geometrical structures with arbitrary variable curvature (Riemannian space and fiber bundles with connection) can be fulfilled within the framework of the so called infinite deformed groups which generalize gauge groups to the case of nontrivial action on the base space of bundles with use of idea of groups deformations.}

Such groups have been constructed in \cite{TMF}. Klein's Erlangen Program was realized for fiber bundles with connection in \cite{7} and for Riemannian space in \cite{8}.

This is important for physics because the widely known gauge approaches to gravity (see, for example, \cite{9}) in fact gives gauge interpretation neither to metric fields nor to frame (tetrad) ones. An interpretation of these as connections in appropriate fibrings has been achieved in way of introduction (explicitly or implicitly) of the assumption about existence of the background flat space (see, for example, \cite{10}). That is unnatural for gravity. The reason for these difficulties lies in the fact that the fiber bundles formalism is appropriate only for the internal symmetry Lie group, which do not act on the space-time manifold. But for the gravity this restriction is obviously meaningless because it does not permit considering gravity as the gauge theory of the translation group.

In this Appendix we also show that\textit{ generalized gauge deformed groups give a group-theoretic description of gauge fields (gravitational field with its metric or frame part similarly to gauge fields of internal symmetry) which is} \textit{alternative for their geometrical interpretation} \cite{11}.

This approach allow to overcome the well known Coleman-Mandula no-go theorem within the framework of generalized gauge deformed groups and gives new possibilities to unification gravity with gauge theory of internal symmetry \cite{12}.

\subsection{\textbf{ Generalized gauge groups }}

Gauge groups of internal symmetry $G^{g} $ are a special case of infinite groups and have simple group structure -- the infinite direct product of the finite-parameter Lie groups $G^{g} =\prod _{x\in M} G$ where product takes on all points $x$ of the space-time manifold $M$. Groups $G^{g} $ act on $M$ trivially: $x'^{\mu } =x^{\mu } $.

For the aim of a generalization of groups $G^{g} $ to the case of nontrivial action on the space-time manifold $M$, let's now consider a Lie group $G_{M}^{} $ with the multiplication law $(\tilde{g}\cdot \tilde{g}')_{}^{\alpha } =\tilde{\varphi }^{\alpha } (\tilde{g},\tilde{g}')$ which act on the space-time manifold $M$ (perhaps inefficiently) according to the formula $x'^{\mu } =\tilde{f}{}^{\mu } (x,\tilde{g})$. The infinite Lie group $\tilde{G}_{M}^{g} $ is parameterized by smooth functions $\tilde{g}^{\alpha } (x)$ which meet the condition $\left|d_{\nu } \tilde{f}^{\mu } (x,\tilde{g}(x))\right|\ne 0$ $\forall x\in M$ (where $d_{\nu } :=d/dx^{\nu } $). For parameters of group $\tilde{G}_{M}^{g} $ we will use the Greek indices from the beginning of the alphabet. The multiplication law in $\tilde{G}_{M}^{g} $ is determined by the formulas:

\begin{equation} \label{eq88} (\tilde{g}\times \tilde{g}')_{}^{\alpha } (x)=\tilde{\varphi }^{\alpha } (\tilde{g}(x),\tilde{g}'(x')),                                                                                         \end{equation}

\begin{equation} \label{eq89} x'^{\mu } =\tilde{f}{}^{\mu } (x,\tilde{g}(x)).                                                                                                                \end{equation}
It is a simple matter to check that these operations truly make $\tilde{G}_{M}^{g} $ a group \cite{TMF}. Formula (\ref{eq89}) sets the action of $\tilde{G}_{M}^{g} $ on $M$. In the case of trivial action of the group $G_{M}^{} $ on $M$, $x'^{\mu } =x^{\mu } $ and $\tilde{G}_{M}^{g} $ becomes the ordinary gauge group $G^{g} =\prod _{x\in M} G$. We name the groups $\tilde{G}_{M}^{g} $ \textit{generalized gauge groups. }

For the clearing of the groups deformations idea we will consider spheres of different radius $R$. All of them have isomorphic isometry groups - groups of rotations $O(3)$. The information about radius of the spheres is in structure constants of groups $O(3)$, which in the certain coordinates may be written as: $F_{12}^{3} =1/R^{2} $, $F_{13}^{2} =-1$, $F_{23}^{1} =1$. Isomorphisms of groups $O(3)$, which change $R$, correspond to deformations.

For gauge groups $\tilde{G}_{M}^{g} $ some isomorphisms also play a role of deformations of space of groups representations, but unlike deformations of finite-parametrical Lie groups such deformations are more substantial, as these allow to independently deform space in its different points.

Let us pass from the group $\tilde{G}_{M}^{g} $ to the group $G_{M}^{g} $ isomorphic to it by the formula $g^{a} (x)=H^{a} (x,\tilde{g}(x))$ (Latin indices assume the same values as the corresponding Greek ones). The smooth functions $H^{a} (x,\tilde{g})$ have the properties:

1) $H^{a} (x,0)=0 \quad \forall x\in M$;

2) $\exists K^{\alpha } (x,g):  \quad K^{\alpha } (x,H(x,\tilde{g}))=\tilde{g}^{\alpha } \quad \forall \tilde{g}\in G, \quad x\in M$.

The group $G_{M}^{g} $ multiplication law is determined by its isomorphism to the group $\tilde{G}_{M}^{g} $ and formulas (\ref{eq88}), (\ref{eq89}):

\begin{equation} \label{eq90} (g*g')_{}^{a} (x)=\varphi _{}^{a} (x,g(x),g'(x')):=H^{a} (x,\tilde{\varphi }(K(x,g(x)),K(x',g'(x')))),
\end{equation}

\begin{equation} \label{eq91} x'^{\mu } =f_{}^{\mu } (x,g(x)):=\tilde{f}{}^{\mu } (x,K(x,g(x)))\,.                                                                               \end{equation}
Formula (\ref{eq91}) sets the action of $G_{M}^{g} $ on $M$.

We name such transformations between the groups $\tilde{G}_{M}^{g} $ and $G_{M}^{g} $ as \textit{deformations} of infinite Lie groups, since (together with changing of the multiplication law) the corresponding deformations of geometric structures of manifolds subjected to group action are directly associated with them. We name the functions $H^{a} (x,\tilde{g})$ \textit{deformation functions}, functions $h(x)_{\alpha }^{a} :=\left. \partial H^{a} (x,\tilde{g})/\partial \tilde{g}^{\alpha } \right|_{\tilde{g}=0} $ \textit{deformation coefficients}, and the groups $G_{M}^{g} $ \textit{infinite (generalized gauge) deformed groups.}

Let us consider expansion

\begin{equation} \label{eq92} \varphi _{}^{a} (x,g,g')=g^{a} +g'^{a} +\gamma (x)^{a} _{bc} g^{b} g'^{c} +{\frac{1}{2}} \rho (x)^{a} _{bcd} g^{d} g'^{b} g'^{c} +... \end{equation}
The functions $\varphi ^{a} $, setting the multiplication law (\ref{eq90}) in the group $G_{M}^{g} $, are explicitly $x$ dependent, so the coefficients of expansion (\ref{eq92}) are $x$ dependent as well. So $x$ dependent became structure coefficients of group $G_{M}^{g} $ (\textit{structure functions }versus structure constants for ordinary Lie groups):

\begin{equation} \label{eq93} F(x)_{bc}^{a} :=\gamma (x)^{a} _{bc} -\gamma (x)^{a} _{cb}  \end{equation}
and coefficients

\begin{equation} \label{eq94} R(x)^{a} _{dbc} :=\rho (x)^{a} _{dbc} -\rho (x)^{a} _{dcb} ,                                                                                       \end{equation}
which we called a \textit{curvature coefficients of the deformed group} $G_{M}^{g} $.

Since $\xi (x)_{a}^{\mu } :=\partial _{a} \left. f_{H}^{\mu } (x,g)\right|_{g=0} =h(x)_{a}^{\alpha } \tilde{\xi }(x)_{\alpha }^{\mu } $, where $\partial _{b} :=\partial /\partial g^{b} $ and $h(x)_{a}^{\alpha } $ is inverse to the $h(x)_{\alpha }^{a} $ matrix, the generators $X_{a} :=\xi (x)_{a}^{\mu } \partial _{\mu } $ ($\partial _{\mu } :=\partial /\partial x^{\mu } $) of the deformed group $G_{M}^{g} $ are expressed through the generators $\tilde{X}_{\alpha } :=\tilde{\xi }(x)_{\alpha }^{\mu } \partial _{\mu } $ of the group $\tilde{G}_{M}^{g} $ by the deformation coefficients: $X_{a} =h(x)_{a}^{\alpha } \tilde{X}_{\alpha } $. So in infinitesimal (algebraic) level, deformation is reduced to independent in every point $x\in M$ nondegenerate liner transformations of generators of the initial Lie group.

\textbf{Theorem A1. }\textit{Commutators of generators of the deformed group }$G_{M}^{g} $\textit{ are liner combinations of generators with structure functions as coefficients:}

\begin{equation} \label{eq95} [X_{a} ,X_{b} ]=F(x)_{ab}^{c} X_{c} \,.                                                                                                        \end{equation}

For generalized gauge nondeformed group $\tilde{G}_{M}^{g} $ we have $[\tilde{X}_{\alpha } ,\tilde{X}_{\beta } ]=\tilde{F}_{\alpha \beta }^{\gamma } \tilde{X}_{\gamma } $, where $\tilde{F}_{\alpha \beta }^{\gamma } $ is structure constants of the initial Lie group $G_{M}^{} $. So for $\tilde{G}_{M}^{g} $  $F_{\alpha \beta }^{\gamma } =\tilde{F}_{\alpha \beta }^{\gamma } $.

\subsection{ \textbf{ Group-theoretic description of connections in fiber bundles and gauge fields of internal symmetry }}

Let $P=M\times V$ be a principal bundle with the base $M$ (space-time) and a structure group $V$ with coordinates $\tilde{\upsilon }^{i} $ and the multiplication law $\, (\tilde{\upsilon }\cdot \tilde{\upsilon }')^{i} =\tilde{\varphi }^{i} (\tilde{\upsilon },\tilde{\upsilon }')$ (indices $i,j,k$). As usually, we define the left $l_{\tilde{\upsilon }} :P=M\times V\to P'=M\times \tilde{\upsilon }^{-1} \cdot V$ and the right $r_{\tilde{\upsilon }} :P=M\times V\to P'=M\times V\cdot \tilde{\upsilon }$ action $V$ on $P$.

Let's consider a group $G_{M} =T_{M} \otimes V$ where $T_{M} $ is the group of space-time translations. The group $G_{M} $ parameterized by the pair $\tilde{t}^{\mu } $, $\tilde{\upsilon }^{i} $, has the multiplication law $(\tilde{g}\cdot \tilde{g}')^{\mu } =\tilde{t}^{\mu } +\tilde{t}'^{\mu } $, $\, (\tilde{g}\cdot \tilde{g}')^{i} =\tilde{\varphi }^{i} (\tilde{\upsilon },\tilde{\upsilon }')$ and act on $M$ inefficiently: $x'^{\mu } =x^{\mu } +\tilde{t}^{\mu } $. One can define the left action of the group $G_{M} $ on the principal bundle $P$: $x'^{\mu } =x^{\mu } +\tilde{t}^{\mu } $, $\upsilon '^{i} =l_{\tilde{\upsilon }}^{i} (\upsilon )$.

The group $\tilde{G}_{M}^{g} $ is parameterized by the functions $\tilde{t}^{\mu } (x)$, $\tilde{\upsilon }^{i} (x)$ which meet the conditions $\left|\delta _{\nu }^{\mu } +\partial _{\nu } \tilde{t}^{\mu } (x)\right|\ne 0$ $\forall x\in M$. The multiplication law in $\tilde{G}_{M}^{g} $ is

\begin{equation} \label{eq96} (\tilde{g}\times \tilde{g}')^{\mu } (x)=\tilde{t}^{\mu } (x)+\tilde{t}'^{\mu } (x')$, \qquad   $\, (\tilde{g}\times \tilde{g}')^{i} (x)=\tilde{\varphi }^{i} (\tilde{\upsilon }(x),\tilde{\upsilon }'(x'))\,,
\end{equation}

\begin{equation} \label{eq97} x'^{\mu } =x^{\mu } +\tilde{t}^{\mu } (x)\,,                                                                                       \end{equation}
where (\ref{eq97}) determines the inefficient action of $\tilde{G}_{M}^{g} $ on $M$ with the kern of inefficiency - gauge group $V^{g} $. The group $\tilde{G}_{M}^{g} $ has the structure $V^{g} \left. \times \right)\, T_{M}^{g} $, which is a semidirect product of the groups $V^{g} $ and $T_{M}^{g} $, act on $P$ as:

\[x'^{\mu } =x^{\mu } +\tilde{t}^{\mu } (x),  \qquad  \upsilon '^{i} =l_{\tilde{\upsilon }(x)}^{i} (\upsilon )\]
and is the group $aut\, P$ of automorphisms of the principal bundle $P$.

Let us deform the group $\tilde{G}_{M}^{g} \to G_{M}^{g} $ by means of deformation functions with additional properties:

3) $H^{\mu } (x,\tilde{t},\tilde{\upsilon })=\tilde{t}^{\mu }  \quad  \forall \; \tilde{t}\in T,\quad \tilde{\upsilon }\in V,\quad x\in M$;

4) $H^{i} (x,0,\tilde{\upsilon })=\tilde{\upsilon }^{i}   \quad \forall \; \tilde{\upsilon }\in V,\quad x\in M$.

The deformed group $G_{M}^{g} $ is parameterized by the functions $t^{\mu } (x)=\tilde{t}^{\mu } (x)$ and \textit{$\upsilon ^{i} (x)=H^{i} (x,\tilde{t}(x),\tilde{\upsilon }(x))$. }Obviously, the group $G_{M}^{g} $, as well as the group $\tilde{G}_{M}^{g} $, has the structure $V^{g} \left. \times \right)\, T_{M}^{g} $ and acts on $P$ as:

\begin{equation} \label{eq98} x'^{\mu } =x^{\mu } +t^{\mu } (x),  \qquad    \upsilon '^{i} =l_{K(x,t(x),\upsilon (x))}^{i} (\upsilon )\,,                                                                     \end{equation}
where functions $K^{i} (x,t(x),\upsilon (x))$ are determined by equation: $K^{i} (x,t(x),\upsilon (x))=\tilde{\upsilon }^{i} (x)$. Properties 3), 4) result in the fact that among deformation coefficients of the group $G_{M}^{g} $, $x$-dependent is only $h(x)_{\mu }^{i} =\left. \partial _{\tilde{\mu }} H^{i} (x,\tilde{t},0)\right|_{\tilde{t}=0} =:-A(x)_{\mu }^{i} $ (where $\partial _{\tilde{\mu }} :=\partial /\partial \tilde{t}^{\mu } $).

Generators of the $G_{M}^{g} $-action on $P$ (\ref{eq98}) are split in the pair

\[X_{\mu } =\partial _{\mu } +A(x)_{\mu }^{i} \tilde{X}_{i} ,   \qquad   X_{i} =\tilde{X}_{i} ,\]
where $\tilde{X}_{i} $ are generators of the left action of the group $V$ on $P$. This result in the natural splitting of tangent spaces $T_{u} $ in any point $u\in P$ to the direct sum $T_{u} =T\tau _{u} \oplus T\upsilon _{u} $ subspaces:

\[T\tau _{u} =\{ t^{\mu } X_{\mu } \} ,   \qquad  T\upsilon _{u} =\{ \upsilon ^{i} X_{i} \} . \]
The distribution $T\tau _{u} $ is invariant with respect to the right action of the group $V$ on $P$, and $T\upsilon _{u} $ is tangent to the fiber. So $T\tau _{u} $ one can treated as horizontal subspaces of the $T_{u} $ and generators $X_{\mu } $ - as covariant derivatives. This set in the principal bundle $P$ connection and deformation coefficients $A(x)_{\mu }^{\, i} $ are the coordinates of the connection form, which on submanifold $M\subset P$ may be written as $\omega ^{i} =A(x)_{\mu }^{\, i} dx^{\mu } $. Necessary condition of existence of group $G_{M}^{g} $ (\ref{eq95}) for generators $X_{\mu } $ gives

\begin{equation} \label{99} [X_{\mu } ,X_{\nu } ]=F(x)_{\mu \nu }^{\, {\kern 1pt} i} X_{i}  \end{equation}
where

\begin{equation} \label{eq100} F(x)_{\mu \nu }^{\, i} =F_{jk}^{\, i} A(x)_{\mu }^{\, j} A(x)_{\nu }^{k} +\partial _{\mu } A(x)_{\nu }^{\, i} -\partial _{\nu } A(x)_{\mu }^{\, i}  \end{equation}
are the structure functions of the group $G_{M}^{g} $ and $F_{jk}^{i} =\tilde{F}_{jk}^{i} $ are the structure constants of the Lie group $V$. Relationship (\ref{eq100}) one can write in the form:

\begin{equation} \label{eq101} d\omega ^{i} =-{\frac{1}{2}} F_{j{\kern 1pt} k}^{{\kern 1pt} i} \omega ^{j} \wedge \omega ^{k} +\Omega ^{i}\,,  \end{equation}
where $\Omega ^{i} ={\frac{1}{2}} F(x)_{\mu \nu }^{\, i} dx^{\mu } \wedge dx^{\nu } $ play the role of the curvature form on submanifold $M$. So equation (\ref{eq101}) is a structural equation for connection, which has been set on principal bundle $P$ by action of group $G_{M}^{g}$.

\textbf{Theorem A2.} \textit{Acting on the principal bundle }$P=M\times V$\textit{ deformed group }$V^{g} \left. \times \right)\, T_{M}^{g} $\textit{ sets on $P$ structure of connection. Any connection on the principal bundle }$P=M\times V$ \textit{may be set in such a way. }

This theorem realizes Klein's Erlangen Program for fiber bundles with connection \cite{7}.

We should emphasize that for the setting of a geometrical structure in $P$ it is enough to consider the infinitesimal action (\ref{eq98}) of the group $G_{M}^{g} $.

The potentials of the gauge field of internal symmetry are identified with deformation coefficients $A(x)_{\mu }^{\, i} $, a strength tensor -- with structure functions $F(x)_{\mu \nu }^{\, i} $ of the group $G_{M}^{g} $ \cite{TMF}. All groups $G_{M}^{g} $ obtained one from another by internal automorphisms, which are generated by the elements $\upsilon (x)\in V^{g} $, describe the same gauge field. These automorphisms lead to gauge transformations for fields $A(x)_{\mu }^{\, i} $ and for infinitesimal $\upsilon ^{i} (x)$ give:

\begin{equation} \label{eq102} A'(x)_{\mu }^{\, i} =A(x)_{\mu }^{\, i} -F_{j{\kern 1pt} k}^{\, i} A(x)_{\mu }^{\, j} \upsilon ^{k} (x)-\partial _{\mu } \upsilon ^{i} (x)\,.                                                                  \end{equation}

\subsection{ \textbf{ Group-theoretic description of Riemannian spaces and gravitational fields }}

The structure of Riemannian space is a special case of structure of affine connection in frame bundle and consequently it can be set by the way described above with the application of the deformed group $SO(n)^{g} \left. \times \right) Diff\, M$. If we force the metricity and non-torsionity conditions, generators of translations $X_{\mu } =\partial _{\mu } +\Gamma (x)_{\mu }^{(mn)} \tilde{S}_{(mn)} $, where $\tilde{S}_{(mn)} $ is generators of group $SO(n)$, become covariant derivatives in Riemannian space. For the setting of Riemannian structure by such means, it is enough to consider the group $SO(n)^{g} \left. \times \right) Diff\, M$ on algebraic level -- on level its generators.

Potentials of a gravitational field in the given approach are represented by the connection coefficients $\Gamma (x)_{\mu }^{(mn)} $ instead of the metrics or verbein fields that would correspond to sense of a gravitational field as a gauge field of translation group which is born by energy-momentum tensor, instead of spin.

Now we will show that the Riemannian structure on $M$ is naturally set also by a narrower group than $SO(n)^{g} \left. \times \right) Diff\, M$, namely, the deformed group of diffeomorphisms $T_{M}^{g} $, though it demands consideration of its action on $M$ up to the second order on parameters.

\textbf{ }Let $G_{M}^{} =T_{M} $ where $T_{M} $ is the group of space-time translations. In this case $(\tilde{t}\cdot \tilde{t}')^{\mu } =\tilde{t}^{\mu } +\tilde{t}'^{\mu } $ and $x'^{\mu } =x^{\mu } +t^{\mu } $. The group $\tilde{T}_{M}^{g} $ is parameterized by the functions $\tilde{t}^{\mu } (x)$, which meet the condition $\left|\delta _{\nu }^{\mu } +\partial _{\nu } \tilde{t}^{\mu } (x)\right|\ne 0$ $\forall x\in M$. The multiplication law in $\tilde{T}_{M}^{g} $ is

\begin{equation} \label{eq103} (\tilde{t}\times \tilde{t}')^{\mu } (x)=\tilde{t}^{\mu } (x)+\tilde{t}'^{\mu } (x')\,,                                                                                         \end{equation}

\begin{equation} \label{eq104} x'^{\mu } =x^{\mu } +\tilde{t}^{\mu } (x)\,,                                                                                                             \end{equation}
where (\ref{eq104}) determines the action of $\tilde{T}_{M}^{g} $ on $M$. The multiplication law indicates that $\tilde{T}_{M}^{g} $ is the group of space-time diffeomorphisms \textit{Diff}$M$ in additive parametrization. Thus, in the approach considered, the \textit{Diff}$M$ group becomes the gauge group of local translations. The generators of the $\tilde{T}_{M}^{g} $ action on $M$ are simply derivatives $\tilde{X}_{\mu } =\partial _{\mu } $ and this fact corresponds to the case of the flat $M$ and the absence of gravitational field.

 Let us deform the group $\tilde{T}_{M}^{g} \to T_{M}^{g} $: $t^{m} (x)=H^{m} (x,\tilde{t}(x))$. The multiplication law in $T_{M}^{g} $ is determined by the formulas:

\begin{equation} \label{eq105} (t*t')^{m} (x)=\varphi ^{m} (x,t(x),t'(x')):=H^{m} (x,K(x,t(x))+K(x',t'(x'))),
\end{equation}

\begin{equation} \label{eq106} x'^{\mu } =f^{\mu } (x,t(x)):=x^{\mu } +K^{\mu } (x,t(x))\,.                                                                               \end{equation}
Formula (\ref{eq106}) sets the action of $T_{M}^{g} $ on $M$.

Let us consider expansion

\begin{equation} \label{eq107} H^{m} (x,\tilde{t})=h(x)_{\mu }^{m} [\tilde{t}^{\mu } +{\frac{1}{2}} \Gamma (x)_{\nu \rho }^{\mu } \tilde{t}^{\nu } \tilde{t}^{\rho } +{\frac{1}{6}} \Delta (x)_{\nu \rho \sigma }^{\mu } \tilde{t}^{\nu } \tilde{t}^{\rho } \tilde{t}^{\sigma } + ...]\,.
\end{equation}
With usage of formula (\ref{eq105}), for coefficients of expansion (\ref{eq92}) we can obtain

\begin{equation} \label{eq108} \gamma ^{m} _{pn} =h_{\mu }^{m} (\Gamma _{p{\kern 1pt} n}^{\mu } +h_{p}^{\nu } \partial _{\nu } h_{n}^{\mu } )\,,                                                                                             \end{equation}

\begin{equation} \label{eq109} \rho ^{m} _{spn} =h_{\mu }^{m} (\Delta _{spn}^{\mu } -\Gamma _{n{\kern 1pt} \sigma }^{\mu } \Gamma _{ps}^{\sigma } -h_{n}^{\nu } \partial _{\nu } \Gamma _{\sigma \rho }^{\mu } h_{p}^{\sigma } h_{{\kern 1pt} s}^{\rho } )\,.
\end{equation}
So formulas (\ref{eq93}), (\ref{eq94}) for structure coefficients and curvature coefficients of deformed group $T_{M}^{g} $ give

\begin{equation} \label{eq110} F_{\mu \nu }^{n} =-(\partial _{\mu } h_{\nu }^{n} -\partial _{\nu } h_{\mu }^{n} )\,,                                                                                                    \end{equation}

\begin{equation} \label{eq111} R^{\mu } _{\nu {\kern 1pt} \kappa \rho } =\partial _{\kappa } \Gamma _{\rho \nu }^{\mu } -\partial _{\rho } \Gamma _{\kappa \nu }^{\mu } +\Gamma _{\kappa \sigma }^{\mu } \Gamma _{\rho \nu }^{\sigma } -\Gamma _{\rho \sigma }^{\mu } \Gamma _{\kappa \nu }^{\sigma } \,.                                                            \end{equation}
In these formulas deformation coefficients $h_{\mu }^{m} $ and $h_{m}^{\mu } $ we use for changing Greek index to Latin (and vice versa).

Formulas (\ref{eq110}) and (\ref{eq111}) show that groups $T_{M}^{g} $ contain information about geometrical structure of space $M$ where they act. The generators $X_{m}^{} =h_{m}^{\nu } \partial _{\nu } $ of the $T_{M}^{g} $-action (\ref{eq106}) on $M$ can be treated as frames (tetrads). Structure functions $F_{\mu \nu }^{n} $ are \textit{anholonomic coefficients} of frames $X_{m}^{} $.

Let us write the multiplication law of the group $T_{M}^{g} $ (\ref{eq105}) for infinitesimal second factor:

\begin{equation} \label{eq112} (t*\tau )^{m} (x)=t^{m} (x)+\lambda (x,t(x))^{m} _{n} \tau ^{n} (x')\,,                                                                          \end{equation}
where $\lambda (x,t)^{m} _{n} :=\partial _{n'} \left. \varphi _{}^{m} (x,t,t')\right|_{t'=0} $. Formula (\ref{eq112}) gives the rule for the addition of vectors, which set in different points $x$ and $x'$ or a \textit{rule of the parallel transport} of a vector field $\tau $ from point $x'$ to point $x$: $\tau _{}^{m} (x)=\lambda (x,t(x))^{m} _{n} \tau ^{n} (x')$, or in coordinate basis $\tau _{}^{\mu } (x)=\partial _{\tilde{\nu }} H^{\mu } (x,\tilde{t})\tau ^{\nu } (x+\tilde{t})$. This formula determines the \textit{covariant derivative}

\begin{equation} \label{eq113} \nabla _{\nu } \tau ^{\mu } (x)=\partial _{\nu } \tau ^{\mu } (x)+\Gamma (x)_{\sigma \nu }^{\mu } \tau ^{\sigma } (x)\,,                                                                                \end{equation}
where functions $\Gamma (x)_{\sigma \nu }^{\mu } $ set the second order of expansion (\ref{eq107}) and play the role of coefficients of an affine connection. They are symmetric on the bottom indexes so torsion equals zero. If $\eta _{mn} $ is a metric of a flat space, in the manifold $M$ we can determine metrics $g_{\mu \nu } =h_{\mu }^{m} h_{\nu }^{n} \eta _{mn} $. In this case generators $X_{m}^{} =h_{m}^{\nu } \partial _{\nu } $ of group $T_{M}^{g} $ form an orthonormal frame (tetrad).

If we force $\gamma _{ksl}^{{\kern 1pt} .} +\gamma _{lsk}^{{\kern 1pt} .} =0$ (lowering indices we perform by metric $\eta _{mn} $), we can show that coefficients $\Gamma _{\mu \nu }^{\rho } $ of expansion (\ref{eq107}) may be written as

\begin{equation} \label{eq114} \Gamma _{\mu \nu }^{\rho } ={\frac{1}{2}} g^{\rho \sigma } (\partial _{\mu } g_{\nu \sigma } +\partial _{\nu } g_{\mu \sigma } -\partial _{\sigma } g_{\mu \nu } )\,.                                                                        \end{equation}
So these coefficients coincide with \textit{Christoffel symbols} $\{ _{\sigma \nu }^{\, \mu } \} $ and curvature coefficients $R^{\mu } _{\nu {\kern 1pt} \kappa \rho } $ of group $T_{M}^{g} $ coincide with the \textit{Riemann curvature tensor}. In this case group $T_{M}^{g} $ is called the group of Riemannian translations \cite{RusMath}.

\textbf{Theorem A3.} \textit{Acting on the manifold $M$ group of Riemannian translations  }$T_{M}^{g} $ \textit{sets on $M$ structure of Riemannian space. Any Riemannian structure on the manifold $M$ may be set in such a way. }

This theorem realizes Klein's Erlangen Program for Riemannian space \cite{8}.

Information about Christoffel symbols is contained in the \textit{second order} of expansion (\ref{eq107}) of deformation functions, and about curvature in functions $\rho ^{m} _{spn} $, which determine \textit{third order} of expansion (\ref{eq92}) in the multiplication law of the group $T_{M}^{g} $. So in this approach we need consider not only infinitesimal (algebraic) level in the group $T_{M}^{g} $ (as in previous section), but higher levels, too.

The gravitational field potentials are identified with deformation coefficients $h(x)_{\mu }^{m} $, strength tensor of the gravitational field -- with structure functions $F(x)_{sn}^{m} $ of the group $T_{M}^{g} $ \cite{TMF}. All groups $T_{M}^{g} $ obtained one from another by internal automorphisms describe the same gravitational field. These automorphisms, which can always be connected with the coordinate transformations on $M$, lead to a general covariance transformation law for fields $h(x)_{\mu }^{m} $ and for infinitesimal $t^{m} (x)$ yield:

\begin{equation} \label{eq115} h'(x)_{\mu }^{m} =h(x)_{\mu }^{m} -F(x)_{sn}^{m} h(x)_{\mu }^{{\kern 1pt} s} t^{n} (x)-\partial _{\mu } t^{m} (x)\,.
\end{equation}

The transformation law (\ref{eq115}) is similar to the transformation law (\ref{eq102}) for gauge fields of internal symmetry and the only difference consists in the replacement of structure constants of finite Lie group by structure functions of the infinite deformed group $T_{M}^{g} $. This fact permits us to interpret the group $T_{M}^{g} $ as the gauge translation group and the vector fields $h_{\mu }^{m} $ as the gauge fields of the translation group.

So we show that infinite deformed (generalized gauge) groups:

a) set on manifolds where they act geometrical structures of Riemannian spaces or fiber bundles with connection with arbitrary variable curvature;

b) give a group-theoretic description of gauge fields -- gravitational field with its frame part and gauge fields of internal symmetry.

\end{document}